\documentclass[prd, floatfix, preprintnumbers, twocolumn, showkeys, 10pt]{revtex4}

\usepackage{amssymb}

\usepackage{epsfig}
\usepackage{epstopdf}

\def\bea{\begin{eqnarray}}
\def\eea{\end{eqnarray}}
\def\be{\begin{equation}}
\def\ee{\end{equation}}
\def\ba{\begin{array}}
\def\ea{\end{array}}
\def\nn{\nonumber}
\newcommand{\mpl}{M_{\rm Pl}}

\renewcommand{\theequation}{\arabic{section}.\arabic{equation}}

\begin{document}

\title{Mass hierarchies and non-decoupling in multi-scalar field dynamics}
\author{Ana Ach\'ucarro$^{a,b}$, Jinn-Ouk Gong$^{a}$, Sjoerd Hardeman$^{a}$, Gonzalo A. Palma$^{c}$, Subodh P. Patil$^{d,e}$}

\affiliation{
$^{a}$Instituut-Lorentz for Theoretical Physics, Universiteit Leiden \mbox{2333 CA Leiden, The Netherlands} \\
$^{b}$Department of Theoretical Physics, University of the Basque Country UPV-EHU,\mbox{P.O. Box 644, 48080 Bilbao, Spain}\\
$^{c}$Physics Department, FCFM, Universidad de Chile \mbox{Blanco Encalada 2008, Santiago, Chile} \\
$^{d}$Laboratoire de Physique Theorique, Ecole Normale Superieure \mbox{24 Rue Lhomond, Paris 75005, France} \\
$^{e}$Centre de Physique Theorique, Ecole Polytechnique and CNRS \mbox{Palaiseau cedex 91128, France} 
}

\begin{abstract}
In this work we study the effects of field space curvature on scalar
field perturbations around an arbitrary background field trajectory
evolving in time. Non-trivial imprints of the `heavy' directions on
the low energy dynamics arise when the vacuum manifold of the
potential does not coincide with the span of geodesics defined by the
sigma model metric of the full theory. When the kinetic energy is
small compared to the potential energy, the field traverses a curve
close to the vacuum manifold of the potential. The curvature of the
path followed by the fields can still have a profound influence on the
perturbations as modes parallel to the trajectory mix with those
normal to it if the trajectory turns sharply enough. We analyze the
dynamical mixing between these non-decoupled degrees of freedom and
deduce its non-trivial contribution to the low energy effective theory
for the light modes. We also discuss the consequences of this mixing
for various scenarios where multiple scalar fields play a vital role,
such as inflation and low-energy compactifications of string theory.
\end{abstract}

\date{\today}
\preprint{CPHT-RR 039.0510, LPTENS-10/20}

\maketitle


\section{Introduction and summary}
\setcounter{equation}{0} A thorough and tractable understanding of
early universe physics through an ultraviolet (UV) complete
description, such as string theory, remains out of our reach for
now. While we do not completely understand the precise features of the
UV completion of the standard model coupled to gravity, or even the
gross structure of such a putative theory, one of its likely generic
consequences is the presence of a large number of scalar fields and
possibly a large number of vacua. The masses of these fields are
typically associated with the internal structure of the theory, for
example, the cutoff scale of the effective action which encapsulates
the UV relevant physics that completes our theory.

However, in general it might be that different fields have different
mass scales associated with them such that a hierarchy appears in the
field content, and we can colloquially speak of `light' and `heavy'
modes. Usually, the heavy fields are assumed to be integrated out
since they are kinematically inaccessible provided that their masses
are larger than the scale of physics of interest. However in reality,
we cannot completely ignore the heavy fields as the conditions
underlying the decoupling theorem of Ref.~\cite{Appelquist:1974tg} can
sometimes be relaxed, for instance through time dependence in the
heavy sector, or through dynamical mixing of heavy and light
sectors. In the context of field theories at a point in field space,
gauge symmetry breaking can lead to a non-decoupling scenario 
(see~\cite{SekharChivukula:2007gi}, and references therein,
for a recent example). In the context of supergravity, decoupling of
heavy fields is still actively being
studied~\cite{Choi:2004sx,deAlwis:2005tg,deAlwis:2005tf,Binetruy:2004hh,Achucarro:2007qa,Achucarro:2008sy,Achucarro:2008fk,BenDayan:2008dv,Gallego:2008qi,
  Brizi:2009nn, Gallego:2009px, Gallego:2011jm}.

\begin{figure}
 \includegraphics[width=0.48\textwidth]{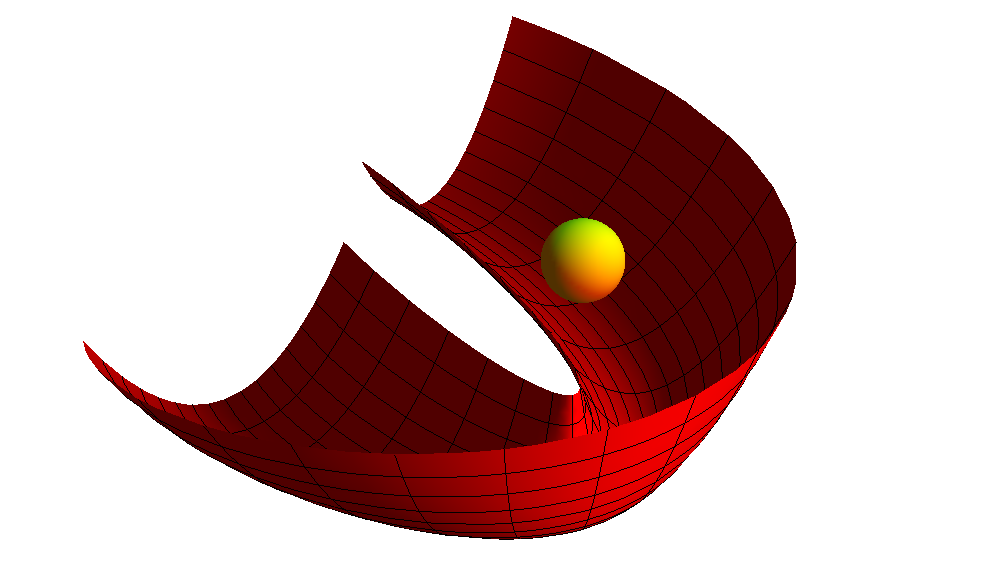}
 \caption{A curved trajectory with a sudden bend, where a scalar
   potential presents a turning flat direction with a massive mode
   remaining perpendicular to it}
 \label{fig:curved_traj}
\end{figure}
In a series of papers, this collaboration will examine the conditions
under which decoupling is relaxed, or fails outright, and explore
quantitatively the consequences of this failure.  The subject of this paper is to study
multi-scalar field theories in Minkowski spacetimes characterized by having a hierarchy of
scales between different families of fields. Our main goal is to
characterize the interaction between light and heavy degrees of
freedom as the background expectation value of the scalar fields
(which we assume to be spatially homogeneous) varies in time,
traversing a curved trajectory in field space. Specifically, we
analyze those situations where the heavy degrees of freedom
correspond to the transverse modes with respect to the path followed
by the background fields. Pictorially, these curved trajectories are
placed in the nearly flat valleys of the vacuum manifold. An example
of such a curved trajectory is schematically shown in
Fig.~\ref{fig:curved_traj} where a scalar potential presents a curved
flat direction with a massive mode remaining perpendicular to it. Such
curvature mixes light and heavy directions along the field trajectory,
possibly non-adiabatically if the field space is `curved' enough.
In the context of inflation, the effects studied in this work lead to
potentially observable effects in the spectrum of primordial
perturbations (see \cite{Achucarro:2010da} and references therein).

In the case of two-field models, we will show that whenever the
trajectory makes a turn in field space such that the light direction
is interchanged by a heavy direction (and vice versa), the low energy
effective action describing the dynamics of the light degrees of
freedom generically picks up a non-trivial correction,
hereby parameterized by $\beta$ and given by
\be
\beta = \ln \left( 1+ \frac{4\dot \phi_0^2}{ \kappa^2 M_H^2} \right) \, ,
\ee
where $\dot \phi_0$ is the velocity of the background scalar field,
$\kappa$ is the radius of curvature characterizing the bending of the
trajectory in field space,  and $M_H$ is the characteristic
mass scale of the heavy modes.  We emphasize that the bending is
measured with respect to the geodesics of the sigma model metric,
whether or not it is canonical; notice that $\kappa$ represents a
distance measured in the geometrical space of scalar fields, and
therefore it has units of mass.

More precisely, at low energies ($k \ll M_H$) the equation of motion describing
the scalar field fluctuation {\em parallel} to the trajectory of
background fields is found to be
\be
\label{effective-eq}
\ddot \varphi - e^{-\beta} \nabla^2 \varphi + M^2_L \, \varphi = 0 \, ,
\ee
with $\nabla^2 \equiv \delta^{ij}\partial_i\partial_j$ being the
spatial Laplacian: see (\ref{low-energy-eff-action}). Thus the kinetic
energy term is modified by a factor of $e^{-\beta}$ in front of the
spatial gradient term, in agreement with
\cite{Tolley:2009fg}. In the context of inflation, this can give rise
to sizable effects that may be observable in the next generation of
the cosmic microwave background observations. For example, for
inflation in supergravity models, the masses of heavy degrees of
freedom during inflation are typically of order $M_H =
\mathcal{O}(H)$, with $H$ being the Hubble parameter during
inflation. This leads to the naive expectation
\be
e^\beta \sim 1+ \epsilon  \frac{M_{\rm Pl}^{2}}{ \kappa^2} \, ,
\ee
where $\epsilon \equiv -\dot{H}/H^2 = (\dot\phi_0/H)^2/(2\mpl^2)$ is
the so-called slow-roll parameter. Then, if the bending of the
trajectory is such that the radius of curvature $\kappa$ becomes much
smaller than $M_{\rm Pl}$ one then is led to $e^\beta
\gg 1$. Such a situation implies strong modifications on
the scale dependence of inflaton fluctuations, contributing
non-trivial features to the power spectrum well within the threshold
of detection. The full analysis valid for inflationary cosmology is,
however, more involved than the naive arguments above as it requires a
detailed account of gravity and careful treatments of the multi-field
dynamics at horizon crossing. For this reason, in the present note we
study the basic features of the effects of the heavy-light field
interactions during turns of the scalar field trajectory in a
Minkowski background, and leave the study of inflationary de Sitter
case for a separate paper~\cite{Achucarro:2010da}. We
stress here that the perspective that follows from our analysis is
that the only relevant criteria for the effects we uncover (such as
the modified speed of sound for the perturbations) is that our system
deviates sufficiently from a geodesic trajectory \footnote{In
  \cite{Tolley:2009fg} \cite{sera}, related effects were uncovered
  that were specifically induced by non-canonical kinetic terms. Our
  perspective is that the latter is more readily thought of as having
  been induced by non-geodesic trajectories in field space. Certainly
  any metric on field space can be brought into canonical form by a
  suitable field redefinition at the expense of introducing
  complicated derivative interactions at higher orders. Therefore the
  criteria must be more general than simply considering non-canonical
kinetic terms (to induce for example, a reduced
  speed of sound for the scalar field perturbations).}.

 Our formalism, and its application to inflation \cite{Achucarro:2010da} builds upon previous
work \cite{GrootNibbelink:2000vx,GrootNibbelink:2001qt,Gordon:2000hv,Nilles:2001fg}
that we extend to be able to account for the effects coming from very sharp
turns and also prolonged turns in which the true ground state deviates
significantly from its single-field approximation. By disregarding the
effects of gravity, we are able to describe in detail the evolution of
the quantum modes in some idealized settings (e.g. turns at a constant 
rate).  We also extend recent results by
\cite{Chen:2009we, Tolley:2009fg}, that we recover in the right limits.
In particular, we are able to clarify the relevance or otherwise of
non-canonical kinetic terms, in particular with respect to a reduced
speed of sound in the effective low energy theory (see also
\cite{sera}). We expect our results to be relevant  for any phenomena where the
variation of vacuum expectation values plays a significant role, such us inflation, phase transitions, soliton interactions and 
soliton-radiation scattering.

In the remainder of this paper, we will first elaborate on the general
framework and study the background equations of motion in
Section~\ref{sec:background}. In Section~\ref{pert-quant}, we will
study the theory of perturbations and the quantization around a
general field trajectory. In Section~\ref{Hierarchies}, we will apply
the theory to the case of a large hierarchy between light and heavy
degrees of freedom that span a curved field space trajectory. To
conclude, we will discuss some applications in
Section~\ref{conclusions}, in particular regarding inflation and the
decoupling of heavy modes in supergravity. Some calculational details
are provided in the appendices.

\section{Geometry of multi-scalar field models}
\label{sec:background}
\setcounter{equation}{0} We are interested in multi-scalar field
theories admitting a low energy effective action consisting of at most
two space-time derivatives. Examples of such theories are low energy
compactifications of string theory and supergravity models, where the
number of scalar degrees of freedom (often referred to as moduli) can
be considerably large. For an arbitrary number $n_{\rm tot}$ of scalar
fields in a Minkowski background, the effective action of such a
theory may be written in the form
\be
\label{initial-action}
S= - \int d^4x \left[ \frac{1}{2}\gamma_{ab}\partial^\mu\phi^a\partial_\mu\phi^b+V(\phi) \right] \, ,
\ee
where $\phi^a$, with $a = 1, \cdots n_{\rm tot}$, is a set of scalar fields and $\gamma_{a b} = \gamma_{a b} (\phi)$ is an arbitrary symmetric matrix restricted to be positive definite.
Any contribution to this action containing higher space-time derivatives will be suppressed by some cutoff energy scale $\Lambda$, which we assume to be much larger than the energy scale of our interest. In low energy string compactifications, such a scale $\Lambda$ corresponds to the compactification scale, which may be close to $M_{\rm Pl}$. However, as mentioned before, we do not consider coupling this theory to gravity, saving this for a separate report~\cite{Achucarro:2010da}.

In order to study the dynamics of the present system it is useful to think of $\gamma_{a b}$ as a metric tensor of some abstract scalar manifold $\mathcal{M}$ of dimension $n_{\rm tot}$. This allows us to consider the definition of several standard geometrical quantities related to $\mathcal{M}$. For instance, the Christoffel connections are given by
\be
 \Gamma^{a}_{b c} = \frac{1}{2} \gamma^{a d} \left( \partial_b \gamma_{d c} + \partial_c \gamma_{b d} - \partial_d \gamma_{b c} \right) \, ,
\ee
where $\gamma^{ab}$ is the inverse metric satisfying $\gamma^{ac}\gamma_{cb} = \delta^a{}_b$, and $\partial_{a}$ denotes a partial derivative with respect to $\phi^a$. This allows us to define covariant derivatives in the usual way as $\nabla_a X_b \equiv \partial_a X_b - \Gamma_{a b}^c X_c$. Also, we may define the associated Riemann tensor as
\be
 \mathcal{R}^a{}_{b c d} \equiv \partial_c \Gamma^a_{b d} - \partial_d \Gamma^a_{b c} + \Gamma^a_{c e} \Gamma^e_{d b} - \Gamma^a_{d e} \Gamma^e_{c b} \, .
 \label{Riemann-def}
\ee
as well as the Ricci tensor $\mathcal{R}_{ab} = \mathcal{R}^{c}{}_{a c b}$ and the Ricci scalar $\mathcal{R} = \gamma^{a b} \mathcal{R}_{a b}$. We should keep in mind that, typically, there will be an energy scale $\Lambda_{\mathcal M}$ associated with the curvature of $\mathcal{M}$, fixing the typical mass scale of the Ricci scalar as $\mathcal{R} \sim \Lambda_{\mathcal M}^{-2}$. In many concrete situations the scale $\Lambda_{\mathcal M}$ corresponds precisely to the aforementioned cutoff scale $\Lambda$. For example, in $N=1$ supergravity realizations of string compactifications, one typically finds that the Ricci scalar of the K\"ahler manifold satisfies $\mathcal{R} \sim M_{\rm Pl}^{-2}$~\cite{GomezReino:2006wv, Covi:2008ea}. In the following, we assume $\Lambda \sim \Lambda_\mathcal{M}$ and do not distinguish them.

Given these geometric quantities, we can write the background equation of motion of the homogeneous background fields $\phi_0^a = \phi_0^a(t)$ from the action (\ref{initial-action}). If $\dot{\phi}_{0a}\dot{\phi}_{0}^{a} \neq 0$, one may think of $\phi_0^a(t)$ as the coordinates parametrizing a trajectory in $\mathcal{M}$, where $t$ is the parameter on the curve describing the expectation value of the scalar fields. Before proceeding with a detailed analysis of these trajectories, let us emphazise that these trajectories may or may not be the geodesics of $\mathcal{M}$, with the details depending on the specific form of the scalar potential $V(\phi)$ and the geometry of $\mathcal{M}$. Varying the action (\ref{initial-action}) with respect to $\phi^a$, the equations of motion of $\phi^a$ are then found to be
\be
 \label{scal-eq-1}
 \frac{D}{dt} \dot \phi_0^a + V^a = 0 \, ,
\ee
where we have introduced the notation $D X^a \equiv d X^a + \Gamma^a_{b c} X^b d \phi_0^c$ and $V^a \equiv  \gamma^{a b} \partial_b V$. Given a certain scalar manifold $\mathcal{M}$ and a scalar potential $V$, these equations can be solved to obtain the trajectory in $\mathcal{M}$ traversed by the background scalar fields. It is convenient to define the rate of variation of the scalar fields $\dot \phi_0$ along the trajectory as
\be
 \dot \phi_0^2 \equiv \gamma_{a b} \dot \phi_0^a  \dot \phi_0^b \, .
\ee
Since $\gamma_{a b}$ is positive definite, $\dot\phi_0^2$ is always greater than or equal to zero. Without loss of generality, in what follows we assume that $\dot{\phi}(t) > 0$ for all $t$. 
Then the unit tangent vector $T^a$ to the trajectory is defined as
\be
T^a \equiv \frac{\dot \phi_0^a}{\dot \phi_0} \, ,
\label{def-T}
\ee
and if the trajectory is curved we can define one of the normal vectors to be
\be
N^a \equiv  s_N (t) \left( \gamma_{bc}\frac{DT^b}{dt}\frac{DT^c}{dt} \right)^{-1/2}\frac{DT^a}{dt} \, ,
\label{def-N}
\ee
where $s_N (t) = \pm 1$ denotes the orientation of $N^a$ with
respect to the vector $D T^a/dt$. That is, if $s_N (t) = +1$ then
$N^a$ is pointing in the same direction as $D T^a/dt$, whereas if $s_N
(t) =-1$ then $N^a$ is pointing in the opposite direction.  Due to the
presence of the square root, it is clear that $N^a$ is only well
defined at intervals where $D T^a/dt \neq 0$. However, since $D
T^a/dt$ may become zero at finite values of $t$, we allow $s_N (t)$
to flip signs each time this happens in such a way that both $N^a$ and $DT^a/dt$
remain a continuous function of $t$. This implies that the sign of $s_N$ may be chosen conventionally at some initial time $t_i$, but from then on it is subject to the equations of motion respected by the background~\footnote{We are assuming here that the background solutions $\phi^a = \phi_0^a(t)$ are analytic functions of time, and therefore we disregard any situation where this procedure cannot be performed.}. 
As we shall see later, in the particular case where $\mathcal{M}$ is two dimensional, the
presence of $s_N(t)$ in (\ref{def-N}) is sufficient for $N^a$ to
have a fixed orientation with respect to $T^a$ (either left-handed or
right-handed).

After some formal manipulations, we can deduce $\kappa$, the radius of curvature of the trajectory followed by the vacuum expectation value $\phi_0^a(t)$, as
\be
 \label{kappa-def}
\frac{1}{ \kappa } = \frac{| V_N |} {\dot \phi_0^2}\, ,
\ee
where $V_N \equiv N^aV_a$. Details of how (\ref{kappa-def}) is derived
can be found in Appendix~\ref{sec:app-geom}. Notice that for $\dot
\phi_0 \neq 0$, the only way of having $V_N = 0$ is by following a
straight curve for which there is no geodesic deviation, $DT^a/dt =
0$. This result follows the basic intuition that, during a turn in the
trajectory, the scalar fields will shift from the minimum of the
potential $V_N = 0$ along the perpendicular direction
$N^a$. Trajectories for which $V_N = 0$ are in fact geodesics, and
therefore the parameter $\kappa$ is a useful measure of the
geodesic deviation caused by the potential $V$. On dimensional
grounds, we should expect $\kappa^{-1} \sim \Lambda_{M}^{-1}$, as the
intrinsic curvature of the manifold $M$ is the result of the embedding
in a yet higher energy completion, and this likely bestows this
manifold with a similar extrinsic curvature scale. This extrinsic
curvature is then necessarily the minimal extrinsic curvature for any
trajectory on $M$. Yet, much more curved trajectories are perfectly
possible~\footnote{For example the degenerate minima offered by the
  $SU(2) \times U(1)$ invariant potential, used to break the
  electroweak symmetry of the standard model, have a radius of
  curvature of order $\kappa \sim M_{\rm H} \ll M_{\rm Pl}$, where
  $M_{\rm H}$ is the Higgs mass.} and we therefore allow a wider range
of values for $\kappa$.


\section{Perturbation theory and its quantization}
\label{pert-quant}
\setcounter{equation}{0}
We now study the evolution of  the scalar field fluctuations around the background time-dependent solution examined in the previous section. We extend the formalism of \cite{GrootNibbelink:2000vx, Nilles:2001fg, GrootNibbelink:2001qt}  to be able to deal with regimes of fast and or continuous turns.  As we shall see, the radius of curvature $\kappa$ couples the perturbations parallel and normal to the motion of the background fields.  Let us start by writing $\phi^a$ including perturbations $\delta \phi^a (t, {\bf x}) \equiv \varphi^a(t, {\bf x}) $ as 
\be
 \label{perturbations-def}
 \phi^a (t, {\bf x})= \phi_0^a(t) + \varphi^a(t, {\bf x}) \, ,
\ee
where $\phi_0^a(t)$ corresponds to the exact solution to the homogeneous equation of motion (\ref{scal-eq-1}). Starting from the action (\ref{initial-action}), one finds the equations of motion for $\varphi^a$ as %
\be
 \label{pert-eq-1}
 \frac{D^2 \varphi^a}{dt^2} - \nabla^2 \varphi^a + C^a_{\phantom{a}b} \varphi^b = 0 \, ,
\ee
where
\be
 C^a_{\phantom{a}b} = \nabla_b V^a - \dot \phi_0^2 \, \mathcal{R}^a_{\phantom{a} c d b} T^c T^d \, .
\ee
It is convenient to rewrite these equations in terms of a new set of perturbations orthogonal to each other. For this we define a complete set of vielbeins $e^I_a$ and work with redefined fields %
\be
 \label{perturbations-def-2}
 v^I = e^I_a \varphi^a \, .
\ee
An example of a possible choice for these vielbeins is a set $\left\{
  e^I_a \right\}$ with $T_a$ and $N_a$ among its elements \cite{Gordon:2000hv}, but we do
not consider this choice until later. Recall that the vielbeins
satisfy the basic relations $\delta_{IJ} e^I_a e^J_b= \gamma_{a b}$
and $\gamma^{a b} e^I_a e^J_b = \delta^{I J}$, which lead to
\bea
e^{I}_a \frac{D}{dt } e^a_J & = & - e^a_J \frac{D}{dt } e^I_a \, ,
\\
e^{a}_I \frac{D}{dt } e_b^I & = & - e_b^I \frac{D}{dt } e_I^a \, .
\eea
In (\ref{perturbations-def-2}) the $\varphi^{a}$ are vectors
with respect to the covariant derivative $D/dt$, while the fields
$v^{I}$ are scalars. Then, it is easy to show that
\bea
 \label{Z-eq-1}
 \dot v^I &=& e^I_a \frac{D \varphi^a}{dt} - Z^I_{\,\,\, J} v^J \, , \\
 \label{Z-eq-2}
 \ddot v^I &=& e^I_a \frac{D^2 \varphi^a}{dt^2} - 2 Z^I_{\,\,\, J} \dot v^J - \left( Z^I_{\,\,\, K} Z^K_{\phantom{a} J} + \dot Z^I_{\,\,\, J} \right) v^J  , \quad
\eea
where we have introduced the antisymmetric matrix $Z^I_{\,\,\, J}$ as
\be
 Z^I_{\,\,\, J} = e^I_a  \frac{D }{dt}  e^a_J \, .
\ee
Notice that $Z^{I}_{\,\,\, J} = \left( e^I_a  \partial_b e^a_J + e^I_a \Gamma^{a}_{b c} e^c_J \right) \dot \phi_0^b = \omega_{b \,\,\, J}^{\,\,\, I} \dot \phi_0^b,$ where $\omega_{b \,\,\, J}^{\,\,\, I}$ are the usual spin connections for non-coordinate basis, therefore $ Z^I_{\,\,\, J} $ can be used to define a covariant derivative acting on the $v^I$-fields as  
\be
 \label{proper-time-der}
 \frac{\mathcal D}{dt} v^{I} \equiv \dot v^I + Z^I_{\,\,\, J} v^J  \, .
\ee
From (\ref{Z-eq-1}) and (\ref{Z-eq-2}), we can see that this new covariant derivative is related to the original covariant derivative as
\bea
 \label{proper-time-der-properties}
 \frac{\mathcal D}{dt} v^{I} & = &  e^I_a \frac{D}{dt} \varphi^{a} \, , \\
 \frac{{\mathcal D}^2}{dt^2} v^{I} & = &  e^I_a \frac{D^2}{dt^2} \varphi^{a} \, .
\eea
One can verify that this derivative is compatible with the Kronecker delta in the sense that $\mathcal D\delta_{IJ}/dt = 0$. Then, it is easy to show that the equation of motion~(\ref{pert-eq-1}) in terms of the new perturbations $v^I$ becomes
\be
 \label{pert-v-eq-2}
 \frac{{\mathcal D}^2}{dt^2} v^{I} - \nabla^2 v^I + C^I_{\,\,J} v^J = 0 \, ,
\ee
where $C^I{}_J \equiv e^I_a e^b_J C^a_{\phantom{a} b}$.

\subsection{Canonical frame}
\label{canonical-frame}
In introducing the vielbeins in the previous section, we have not specified any alignment of the moving frame. In fact, given an arbitrary frame $e^{I}_a$, it is always possible to find a `canonical' frame where the scalar field perturbations acquire canonical kinetic terms in the action. To find it, let us introduce a new set of fields $u^I$ defined from the original ones $v^I$ in the following way:
\be
 \label{rel-u-v-1}
 v^I (t) = R^I_{\phantom{I} J} (t , t_0 ) u^J (t) \, .
\ee
Here, $R^I_{\phantom{I} J}(t , t_0 )$ is a matrix satisfying the first order differential equation
\be
 \label{R-def}
 \frac{d}{d t} R^{I}_{\phantom{I} J} = - Z^{I}_{\phantom{I} K} R^{K}_{\phantom{K} J} \, ,
\ee
with the boundary condition $R^I_{\phantom{I} J}(t_0 , t_0 ) = \delta^I{}_J$. By defining a new matrix $S^{I}_{\phantom{I} J}$ to be the inverse of  $R^{I}_{\phantom{I} J}$, namely $S^{I}_{\phantom{I} K} R^{K}_{\phantom{K} J} = R^{I}_{\phantom{I} K} S^{K}_{\phantom{K} J} = \delta^{I}{}_{J}$, it is not difficult to see that $S^{I}_{\phantom{I} J}$ satisfies 
\be
 \label{S-def}
 \frac{d}{d t} S_{I}^{\phantom{I} J} = - Z^{J}_{\phantom{J} K} S_{I}^{\phantom{I} K} \, ,
\ee
where we used the fact that $Z$ is antisymmetric in its indices $Z_{I J} = - Z_{J I}$. Since both solutions to (\ref{R-def}) and (\ref{S-def}) are unique, the previous equation tells us that $S_{IJ} = R_{J I}$. Thus, $S$ corresponds to $R^T$, the transpose of $R$. This means that for a fixed time $t$, $R_{IJ}(t, t_0)$ is an element of the orthogonal group O$(n_{\rm tot})$, the group of matrices $R$ satisfying $R R^T = 1$. The solution of (\ref{R-def}) is well known, and may be symbolically written as
\bea
 R (t , t_0) &=& 1 + \sum_{n=1}^{\infty} \frac{(-1)^n}{n!} \!\! \int_{t_0}^t  \mathbb{T} \left[Z (t_1) \cdots Z(t_n) \right] d^n t \nonumber \\
 &=& \mathbb{T} \exp \left[ -\int_{t_0}^t \!\!\! dt \, Z(t) \right] \, ,
\eea
where $\mathbb{T}$ stands for the time ordering symbol. That is: $\mathbb{T} [Z (t_1) Z (t_2) \cdots Z(t_n)]$ corresponds to the product of $n$ matrices $Z (t_i)$ for which $t_1 \ge t_2 \ge \cdots \ge t_n$. Coming back to the $u^I$-fields, it is possible to see now that, by virtue of (\ref{R-def}) one has
\bea
 \frac{\mathcal{D} v^I}{d t} & = & R^{I}_{\phantom{I} J} \frac{d u^{J}}{d t} \, , \\
 \frac{\mathcal{D}^2 v^I}{d t^2} & = & R^{I}_{\phantom{I} J} \frac{d^2 u^{J}}{d t ^2} \, .
\eea
Inserting these relations back into the equation of motion (\ref{pert-v-eq-2}), we obtain the equation of motion for the $u^I$-fields as
\be
 \label{eom-u-cov-1}
 \frac{d^2u^I}{dt^2} - \nabla^2 u^I + \left[  R^T(t) \, C \, R(t) \right]^{I}_{\phantom{I} J} u^J = 0 \, .
\ee
This equation of motion can be derived from an action
\bea
 S &=& \frac{1}{2} \int d^4x \big\{ \dot u^I \dot u_I - \nabla u^I \cdot \nabla u_I \nn\\
 && \qquad - \left[  R^T (t) C R(t) \right]_{I J} u^I  u^J \big\} \, .   \label{canonical-action}
\eea
Thus, we see that the $u^I$-fields correspond to the canonical fields in the usual sense. To finish, let us notice that by construction, at the initial time $t_0$, the canonical fields $u^I$ and the original fields $v^I$ coincide, $u^I (t_0) = v^I (t_0)$. Obviously, it is always possible to redefine a new set of canonical fields by performing an orthogonal transformation.

\subsection{Quantization of the system}
Having the canonical frame at hand, we may now quantize the system in
the usual way. Starting from the action (\ref{canonical-action}), the canonical momentum conjugate to $u^I$ is given by $\Pi_u^I
= du^I/dt$. To quantize the system, we then demand this pair to
satisfy the canonical commutation relations
\bea
 \left[ u^I({\bf x},t),\Pi_u^J ({\bf y},t) \right] &=& i\delta^{IJ}\delta^{(3)}({\bf x}-{\bf y}) \, ,
\eea
with all other commutators vanishing. With the help of the $R$ transformation introduced in (\ref{rel-u-v-1}), we can define commutation relations which are valid in an arbitrary moving frame. More precisely, we are free to define a new canonical pair $v^I$ and $\Pi^I_v$ given by
\bea
 v^I &=& R^{I}_{\phantom{I} J} u^J \, , \\
 \Pi_{v }^{I} & \equiv & \frac{\mathcal D}{d t} v^I = R^{I}_{\phantom{I} J}(t,t_0) \Pi_{u}^{J} \, .
\eea
This pair is found to satisfy similar commutation relations \cite{GrootNibbelink:2001qt}
\bea
 \label{commutation-1}
 \left[ v^I ({\bf x},t),\Pi_v^J({ \bf y},t) \right] = i\delta^{IJ}\delta^{(3)}({\bf x}-{\bf y}) \, .
\eea
It is in fact possible to obtain an explicit expression for $v^I({\bf x},t)$ in terms of creation and annihilation operators. For this, let us write $v^I({\bf x} , t)$ as a sum of Fourier modes %
\bea
 \label{v-fourier}
 v^I ({\bf x}, t )  &=&  \frac{1}{(2 \pi)^{3/2}} \sum_{\alpha }  \int d^3 k  \big[  e^{ i {\bf  k} \cdot {\bf x} } \,  v_{\alpha }^I (k, t) \, a_\alpha ({\bf k}) \nn \\ 
 && \qquad + e^{-i {\bf  k} \cdot {\bf x} } \,  v_{\alpha }^{I *}(k, t) \, a^{\dag}_\alpha ({\bf k}) \big] \, ,
\eea
where we have anticipated the need of expressing $v^I ({\bf x}, t )$ as a linear combination of $n_{\rm tot}$ time-independent creation and annihilation operators $a^{\dag}_\alpha ({\bf k})$ and $a_\alpha ({\bf k})$ respectively, with $\alpha = 1, \cdots , n_{\rm tot}$. These operators are required to satisfy the usual commutation relations %
\be
 \left[ a_\alpha ({\bf k}) , a_\beta^{\dag}({\bf k'}) \right] = \delta_{\alpha \beta}  \delta^{(3)} ({\bf k} - {\bf k'}) \, ,
\ee
otherwise zero. Since the operators $a^{\dag}_\alpha ({\bf k})$, for different $\alpha$, are taken to be linearly independent, the time-dependent coefficients $v_{\alpha}^I(k,t)$ in (\ref{v-fourier}) must satisfy $n_{\rm tot}$ independent equations (one for each value of $\alpha$) given by
\be
 \frac{\mathcal{D}^2}{d t^2} v_{\alpha}^{I} (t , k) + k^2 v_{\alpha}^{I} (k , t) + C^{I}_{\phantom{I} J} v_{\alpha}^{J} (k , t ) = 0 \, ,
\ee
which follows from (\ref{pert-v-eq-2}). Of course, there must exist $n_{\rm tot}$ independent solutions $v_{\alpha }^I (k, t)$ to this equation. 
\begin{widetext}
For definiteness, at a given initial time $t = t_0$ we are allowed to choose each mode to satisfy the following orthogonal initial conditions %
\be
 v_{1}^I  = \left(
  \begin{array}{c}
    1 \\0 \\ \vdots \\0
  \end{array}
 \right) v_1(k) , \quad v_{2}^I  = \left(
  \begin{array}{c}
    0 \\1 \\ \vdots \\0
  \end{array}
 \right) v_2(k), \quad \cdots \quad v_{n_{\rm tot}}^I  = \left(
  \begin{array}{c}
    0 \\0 \\ \vdots \\1
  \end{array}
 \right) v_{n_{\rm tot}}(k) , \label{initial-1}
\ee
and initial momenta
\be
 \frac{ \mathcal D v_{1}^{I} }{dt}  = \left(
  \begin{array}{c}
    1 \\0 \\ \vdots \\0
  \end{array}
 \right) \pi_1(k) , \quad \frac{ \mathcal D v_{2}^{I} }{dt} = \left(
  \begin{array}{c}
    0 \\1 \\ \vdots \\0
  \end{array}
 \right) \pi_2(k), \quad \cdots \quad \frac{ \mathcal D v_{n_{\rm tot}}^{I} }{dt} = \left(
  \begin{array}{c}
    0 \\0 \\ \vdots \\1
  \end{array}
 \right) \pi_{n_{\rm tot}}(k) , \label{initial-2}
\ee
\end{widetext}
where each unit vector refers to a direction in the tangent space
labelled by the $I$-index, and $v_{\alpha}(k)$ and $\pi_{\alpha}(k)$
are the factors defining the amplitude of the initial
conditions. Since the operator $\mathcal{D}/dt = \partial/\partial t +
Z$ mixes different directions in the $v^I$-field space and since in
general the time-dependent matrix $C_{IJ}$ is non-diagonal, then the
mode solutions $v_{\alpha }^I (k, t)$ satisfying the initial
conditions (\ref{initial-1}) and (\ref{initial-2}) will not
necessarily remain pointing in the same direction nor will they remain
orthogonal at an arbitrary time $t \neq t_0$. Therefore, each mode
solution labelled by $\alpha$ will provide a linearly independent
vector which is allowed to vary its direction in the field space
labelled by the $I$-index, which is why one needs to
distinguish between the two set of indices. 
In other words the $I$-index labels the system of coupled oscillators while the $\alpha$-index labels the modes.

In Appendix~\ref{sec:app-init-cond} we show that it is always possible
to choose initial conditions for $v_{\alpha}^{I} ( k,t)$ and
$\mathcal{D}v_{\alpha}^{I} (k,t)/dt$ such that the commutation
relations (\ref{commutation-1}) are ensured. The choice of the initial
conditions in (\ref{initial-1}) and (\ref{initial-2}) is a convenient
one, as it is not difficult to verify that with this choice and the
requirement on $ v_\alpha(k)$ and $\pi_{\alpha}(k)$ such that %
\be\label{eq:intialrequirement} v_\alpha(k) \pi_{\alpha}^*(k) -
v_\alpha^* (k) \pi_{\alpha}(k) = i , \ee and the commutation relations
(\ref{commutation-1}) are automatically satisfied for all times. We
should emphazise however that, as discussed in the appendix, this is
not the unique choice for initial conditions. In general, any choice
satisfying conditions (\ref{quantum-conditions1}) and
(\ref{quantum-conditions2}) deduced and discussed in
Appendix~\ref{sec:app-init-cond} will do just fine.

\section{Dynamics in the presence of mass hierarchies}
\label{Hierarchies}
\setcounter{equation}{0}
The main quantity determining the dynamics of the present system is the scalar potential $V(\phi)$. Since we are interested in studying the dynamics of multi-scalar field theories in Minkowski space-time, we will assume that it is positive definite, $V (\phi) \ge 0$. From the potential one can define the mass matrix $M^2_{a b}$ associated to the scalar fluctuations around a given vacuum expectation value $\langle \phi^a \rangle = \phi^a_0$ as
\be
 M^2_{a b} (\phi_0) \equiv \nabla_a \nabla_b V \big|_{\phi = \phi_0} \, \label{Mass-matrix} .
\ee
In general this definition renders a non-diagonal mass matrix, yet it is always possible to find a `local' frame in which it becomes diagonal, where the entries are given by the eigenvalues $m_a^2$. Now, we take into account the existence of hierarchies among different families of scalar fields, and specifically consider two families, herein referred to as heavy and light fields which are characterized by 
\be
 \label{hierarchy}
 m^2_{H} \gg m^2_{L} \, .
\ee
In the particular case where the vacuum expectation value of the scalar fields remains constant $\dot \phi^a_0 = 0$, it is well understood that the heavy fields can be systematically integrated out, providing corrections of $\mathcal{O}(k^2/m^2_H)$ with $k$ being the energy scale of interest to the low energy effective Lagrangian describing the remaining light degrees of freedom~\cite{Appelquist:1974tg}. If however the vacuum expectation value $\phi^a_0$ is allowed to vary with time, new effects start occurring which can be significant at low energies. Naively speaking, given the mass matrix (\ref{Mass-matrix}) one would say that a hierarchy between the directions $T^a$ and $N^a$ is present if
\bea
 T^a T^b M^2_{a b} & \sim & m_{L}^2 \, , \label{hierarchy-naive-1} \\
 N^a N^b M^2_{a b} & \sim & m_{H}^2 \, . \label{hierarchy-naive-2} 
\eea
However, the flatness of the potential along the $T^a$ direction is rather given by the following condition
\be
T^a \nabla_a (T^b \nabla_b V)  \sim  m_{L}^2.
\ee
After recalling the definitions (\ref{def-N}) and (\ref{kappa-def}) and using the geometrical identity $T^a \nabla_a = \dot \phi^{-1} D/dt$ along the trajectory, we arrive to the more refined requirements
\bea
  T^a T^b M^2_{a b} - \frac{V_N^2}{\dot \phi^2}  & \sim & m_{L}^2 \, , \label{hierarchy-1} \\
 N^a N^b M^2_{a b} & \sim & m_{H}^2 \, . \label{hierarchy-2} 
\eea
We therefore focus on scalar potentials $V(\phi)$ for which a hierarchy of the type (\ref{hierarchy-1}) and (\ref{hierarchy-2}) is present.
Since $M^2_{ab}$ is in general non-diagonal, for consistency we take $T^a N^b M^2_{a b}$ to be at most of $\mathcal{O}( m_{L} m_H)$. Such trajectories are generic in the following sense: for arbitrary initial conditions, the background field $\phi_0^a$ typically will start evolving to the minimum of the potential $V(\phi)$ by first quickly minimizing the heavy directions. Then the light modes evolve to their minimum  much more slowly.

We will continue the present analysis systematically by splitting the potential into two parts,
\be
 V (\phi) = V_*(\phi) + \delta V(\phi) \, .
 \label{splitting-V}
\ee
Here, $V_* (\phi) \ge 0$ is the zeroth-order positive definite
potential characterized by containing exactly flat directions, and
$\delta V(\phi)$ is a correction which breaks this flatness. Such a type of splitting happens, for
instance, in the moduli sector of many low energy string
compactifications, where $V_{*}$ appear as a consequence of
fluxes~\cite{Giddings:2001yu} and $\delta V(\phi)$ arguably from
non-perturbative effects~\cite{Kachru:2003aw}.  

By construction, $V_*(\phi)$ contains all the information regarding
the heavy directions. Therefore, the mass matrix $M_{* \, ab}^2$
obtained out of $V_*(\phi)$ presents eigenvalues which are either zero
or $\mathcal{O}(m_H^2)$, and the light masses appear only
after including the correction $\delta V(\phi)$. We thus require the
second derivatives of $\delta V(\phi)$ to be at most
$\mathcal{O}(m_L^2)$. It should be clear that such a splitting is not
unique, as it is always possible to redefine both contributions while keeping the property $M_{* \, ab}^2 \sim m_H^2$. 
In Appendix \ref{appendix-zero-order} we study in detail the dynamics offered by the zeroth order contribution $V_*(\phi)$ to the potential
and show how the effects of geodesic deviations take place on the evolution of background fields for slow turns.
In what follows we use these criteria to disentangle light and heavy physics, leading us to an effective description of 
this class of system at low energy.

\subsection{Two-field models}
For theories with two scalar fields we can always choose the set of vielbeins $e^I_a$ to consists in the following pair:
\bea
 e_1^a & = & e_T^a = T^a  , \\
 e_2^a & = & e_N^a =  N^a  .
\eea
This is in fact allowed by the presence of $s_N(t)$ in definition (\ref{def-N}). A concrete choice for $T^a$ and $N^a$ with the properties implied by (\ref{def-N}) is given by
\bea
T^a &=&  \, \frac{1}{\dot \phi_0} \left( \dot \phi^1 , \dot \phi^2 \right), \label{oriented-T} \\
N^a &=& \, \frac{1}{\dot \phi_0 \sqrt{ \gamma} } \left(- \gamma_{22} \dot \phi^2 - \gamma_{12} \dot \phi^1 , \gamma_{11} \dot \phi^1+ \gamma_{21} \dot \phi^2 \right), \qquad  \label{oriented-N}
\eea
where $\gamma = \gamma_{11} \gamma_{22} - \gamma_{12} \gamma_{21}$ is the determinant of $\gamma_{ab}$.
With this choice  we can write $v^T \equiv T_a\varphi^a$ and $v^N \equiv N_a\varphi^a$ from (\ref{perturbations-def-2}), which denote the perturbations parallel and normal to the background trajectory, respectively. In the case where $\mathcal M$ is two-dimensional, these mutually orthogonal vectors are enough to span all of space. Therefore, the two unit vectors satisfy the relations
\bea
 \frac{D T^a}{d t} & = &  - \, \zeta \, N^a \, , \\
 \frac{D N^a}{d t} & = &  \zeta \, T^a \, ,
\eea
where we have defined the useful parameter:
\be
\zeta \equiv \frac{V_N}{ \dot \phi_0} .
\ee
From eq.~(\ref{kappa-def}) we also have $| \zeta | = \dot \phi_0 / \kappa$. In terms of the formalism of the previous sections, we may write $Z_{T N} = - Z_{N T} =   \zeta$. Further, the entries of the symmetric tensor $C_{IJ} = e_{I}^a e_{J}^b C_{ab}$ defined in (\ref{pert-v-eq-2}) are given by %
\bea
 C_{TT} & = & T^a T^b \nabla_{a} V_{b} \, , \\
 C_{TN} & = & T^a N^b \nabla_{a} V_{b} \, , \\
 C_{NN} & = & N^a N^b \nabla_{a} V_{b}  + \frac{\dot \phi_0^2}{2}  \mathcal{R} \, ,
\eea
where $\mathcal{R} = \gamma^{a b} \mathcal{R}^{c}_{\,\,\, a c b} =  2 \mathcal{R}^{T}_{\,\,\, N T N}$ is the Ricci scalar.\footnote{Since $\mathcal M$ is two dimensional,  $\mathcal{R}^{T}_{\,\,\, N T N} = \mathcal{R}/2$ is the only non-vanishing component of the Riemann tensor.} Then, by noticing that $T^a T^b \nabla_{a} V_{b} = T^a \nabla_a \left(T^b V_b\right) - \left( T^a \nabla_a T^b \right) V_b$ and using the fact $T^a \nabla_a \equiv \nabla_\phi = \dot \phi_0^{-1} D/dt$, we may rewrite
\bea
 C_{TT} & = & \nabla_\phi V_\phi + \zeta^2 \, , \label{C-TT} \\
 C_{TN} & = & \dot\zeta - \frac{2V_\phi}{\kappa} \, .
\eea
The remaining component $C_{NN}$ cannot be deduced in this way, as it depends on the second variation of $V$ away from the trajectory. From our discussion regarding equations
(\ref{hierarchy-1}) and (\ref{hierarchy-2}), we demand $\nabla_\phi V_\phi  \sim m_L^2$ and $C_{NN} \sim m_H^2$ such that $m_L^2 \ll m_H^2$.

Inserting the previous expressions back into (\ref{pert-v-eq-2}), the set of equations of motion for the pair of perturbations $v^T$ and $v^N$ is found to be
\bea
 \label{eq-psi-inte}
\Box v^T  - 2 \zeta \dot v^N -  2 \dot \zeta v^N   - \nabla_{\phi} V_{\phi}   v^T + 2 \frac{V_{\phi}}{\kappa} v^N & = & 0  , \qquad  \\
 \label{eq-psi-inte-2}
\Box v^N + 2 \zeta \dot v^T   - M^2 v^N  + 2 \frac{V_{\phi}}{\kappa} v^T & = & 0  ,
\eea
where $\Box = - \partial_t^2 + \nabla^2 $ and $M^2 = C_{NN} - \zeta^2$. It is interesting to see that the contribution $\zeta^2$ to (\ref{C-TT}) has cancelled out in the equations of motion.  The rotation matrix $R^I_{\,\,\, J}$ connecting the perturbations $v^I$ with the canonical counterparts $u^I$ is easily found to be
\bea
 R^I_{\,\,\, J} & = & \left(\begin{array}{cc} \cos \theta(t) & - \sin \theta(t) \\ \sin \theta(t) & \cos \theta(t) \end{array}\right) \, , \\
 \theta(t) & = & \int_{t} ~ dt' \, \zeta (t') \, .
\eea
The convenience of staying in the frame where $e_T^a = T^a$ and $e_N^a =  N^a$ is that the matrix $C_{IJ}$ has elements with a well defined physical meaning.

\subsection{An exact analytic solution : Constant radius of
  curvature} \label{section: constant radius of curvature} To gain
some insight into the dynamics behind these equations, let us consider
the particular case where $V_{\phi} = T^a V_a = 0$ and $\nabla_{\phi}
V_{\phi} = 0$. This is the situation in which the background solution
consists of a trajectory in field space crossing an exactly flat
valley within the landscape. As $V_{\phi} = 0$ requires $\ddot \phi_0
= 0$, we see that $\dot \phi_0$ becomes a constant of
motion. Additionally, let us assume that the radius of curvature
$\kappa$ remains constant, and that the mass matrix $M^2 = C_{NN} -
\zeta^2$ is also constant~\footnote{Notice that in the particular case
  where $\mathcal M$ is flat and with a trivial topology, these
  conditions would correspond to an exact circular curve, such as the
  one that would happen at the bottom of the `Mexican hat'
  potential.}. Under these conditions $\zeta$ is a constant and one
has $C_{TT} = \zeta^2$ and $C_{TN} = 0$. Then, the equations of motion
for the perturbations become
\bea
 \ddot v^T -\nabla^2 v^T  + 2 \zeta \dot v^N &  = & 0 \, , \label{eq-psi-inte-alt} \\
 \ddot v^N  -\nabla^2 v^N - 2 \zeta \dot v^T   + M^2 v^N & = & 0  \label{eq-psi-inte-alt-2} \, .
\eea
We can solve and quantize these perturbations by following the
procedure described in Section~\ref{pert-quant}. First, the
mode solutions $v^I_{\alpha} (k)$ must satisfy
\bea
 \ddot v^T_{\alpha} + 2 \zeta \dot v^N_{\alpha} + k^2  v^T_{\alpha} & = & 0 \, , \label{eq-flat-v-T} 
 \\
 \ddot v^N_{\alpha}  - 2 \zeta \dot v^T_{\alpha} + \left( M^2 + k^2 \right) v^N_{\alpha} & = & 0 \, . \label{eq-flat-v-N}
\eea
To obtain the mode solutions let us try the ansatz
\bea
 v^T_{\alpha} (k,t) =v_{\alpha}^T (k)  e^{- i \omega_{\alpha} t}, \\
 v^N_{\alpha} (k,t) =v_{\alpha}^N (k)  e^{- i \omega_{\alpha} t},
\eea
where $\omega_{\alpha} \ge 0$ ($\alpha = 1,2$) corresponds to a set of frequencies to be deduced shortly. Notice that the associated operators $a_{\alpha}^{\dag}(\bf k)$ and $a_{\alpha}(\bf k)$ create and annihilate quanta characterized by the frequency $\omega_{\alpha}$ and momentum ${\bf k}$. Before proceeding, it is already clear that due to the mass hierarchy we will obtain a hierarchy for the frequencies. This frequency hierarchy precisely dictates what is meant by heavy and light, and therefore the fields $v^T$ and $v^N$ associated to directions of the trajectory in field space are combinations of both light and heavy modes. With the former ansatz, the equations of motion take the form
\bea
 \left( k^2- \omega_{\alpha}^2 \right) v^T_{\alpha} (k)  - 2 i \omega_{\alpha} \zeta v^N_{\alpha} (k) &    = & 0 \, , \label{eq-psi-inte-9} \\
 \left( M^2 + k^2 - \omega_{\alpha}^2 \right) v^N_{\alpha} (k)   + 2 i \omega_{\alpha} \zeta v^T_{\alpha} (k) & = & 0  \label{eq-psi-inte-10} \, .
\eea
Combining them one finds the equation determining the values of $\omega_{\alpha}$ as
\be
 \left( k^2 - \omega_{\alpha}^2 \right) \left( M^2 +k^2 - \omega_{\alpha}^2 \right) = 4 \zeta^2 \omega_{\alpha}^2 \, .
\ee
The solutions to this equation are
\bea
 \omega^{2}_{\pm} &=& \frac{1}{2} \bigg[ \left( M^2 + 2 k^2 + 4 \zeta^2 \right)  \nn
 \\ && \pm \sqrt{\left( M^2 + 2 k^2 + 4 \zeta^2 \right)^2 - 4 k^2 \left( M^2 + k^2 \right)}  \bigg]  . \,\,\, \qquad
\eea
On the other hand, the coefficients $v_{\alpha}^T (k)$ and
$v_{\alpha}^N (k)$ must be such that relations (\ref{cond-v-1}) and
(\ref{cond-v-2}) are satisfied. After straightforward algebra, it is
possible to show that these coefficients are given by
\bea
 |v^T_{-}(k)|^2 &=& \frac{\left(\omega_{+}^2 - k^2\right) \omega_{-}}{2 k^2 \left(\omega_{+}^2 - \omega_{-}^2\right)} \, ,  
 \\
 |v^N_{-}(k)|^2 &=&  \frac{ \left(\omega_{+}^2 - M^2 - k^2 \right) \omega_{-}}{2 \left(M^2 + k^2\right) \left(\omega_{+}^2 - \omega_{-}^2\right)} \, , 
 \\
 |v^T_{+}(k)|^2 &=& \frac{\left(k^2 - \omega_{-}^2 \right) \omega_{+}}{2 k^2 \left(\omega_{+}^2 - \omega_{-}^2\right)} \, , 
 \\
 |v^N_{+}(k)|^2 &=&  \frac{ \left(M^2 + k^2 - \omega_{-}^2 \right) \omega_{+}}{2 \left(M^2 + k^2\right) \left(\omega_{+}^2 - \omega_{-}^2\right)} \, .
\eea
In the low energy regime $k^2 \ll M^2$ we can in fact expand all the relevant quantities in powers of $k^2$. One finds, up to leading order in $k^2/M^2$,
\bea
 \label{omega-minus}
 \omega^2_{-} &=& k^2  \left(1 - \frac{4 \zeta^2}{M^2+ 4 \zeta^2} \right) , \\
 \omega^2_{+} &=& M^2+ 4 \zeta^2+  k^2  \left(1 + \frac{4 \zeta^2}{M^2+ 4 \zeta^2} \right) , \quad \\
  \label{expansion-v-1}
 |v^T_{-}(k)|^2 &=& \frac{M}{2k \sqrt{M^2 + 4 \zeta^2}} , \\
 \label{expansion-v-2}
 |v^N_{-}(k)|^2 &=& \frac{2 \zeta^2 k}{ M (M^2 + 4 \zeta^2)^{3/2}} , \\
\label{expansion-v-3}
 |v^T_{+}(k)|^2 &=&   \frac{2 \zeta^2 }{  (M^2 + 4 \zeta^2)^{3/2}} , \\
\label{expansion-v-4}
 |v^N_{+}(k)|^2 &=& \frac{1}{2\sqrt{M^2 + 4 \zeta^2}} .
\eea
Thus we see that in the particular case where $\zeta = 0$ at all times
(a straight trajectory) one has $|v^N_{-}(k)|^2 = 0$ and
$|v^T_{+}(k)|^2=0$ and one recovers the standard results describing
the quantization of a massless scalar field $v^T_{-}$ and a massive
scalar field $v^N_+$ of mass $M$. Observe that in this case it was not
necessary to choose eqs. (\ref{initial-1}), (\ref{initial-2}) and
(\ref{eq:intialrequirement}) as initial conditions to ensure the
quantization of the system. Note
also that the results hold regardless of whether the sigma model
metric is canonical or not.

\subsection{The general case: low energy effective theory in the
  presence of a very heavy mode}
Although in general it is not possible to solve (\ref{eq-psi-inte})
and (\ref{eq-psi-inte-2}) analytically, we may integrate the heavy
mode to deduce a reliable low energy effective theory describing the
light degree of freedom parallel to the trajectory as long as $k \ll
M$. In the example of the previous section, heavy and light modes were identified with 
the set of frequencies $\omega_+$ and $\omega_-$, and found to be closely related to the 
respective directions $N$ and $T$ in field space. Following that guideline, here we  adopt the notation
$\alpha = L, H$ and focus on the light mode $v_{L}^I$, which here we
express as %
\be v^I_L \to \left(\begin{array}{c} v^T_L \\ v^N_L \end{array}\right)
\equiv \left(\begin{array}{c} \psi \\ \chi \end{array}\right) \, ,
 \label{v-N-0}
\ee
where $\chi$ is a contribution satisfying $| \ddot \chi | \ll M^2 | \chi|$, that is, its time variation is much slower than the time scale $M^{-1}$ characterizing the heavy mode. Then, inserting (\ref{v-N-0}) back into the second equation of motion~(\ref{eq-psi-inte-2}) and keeping the leading term in $\chi$, we obtain the result
\be
 \chi = \frac{2 \zeta}{M^2}  \dot \psi  + 2 \frac{V_{\phi}}{M^2 \kappa} \psi \, .  \label{v-N-chi}
\ee
Of course, we have to verify that  $| \ddot \chi | \ll M^2 | \chi|$ is a good ansatz for the solution. Inserting (\ref{v-N-chi}) back into the first equation of motion~(\ref{eq-psi-inte}) we obtain
\bea
 \ddot \psi + 4 \frac{d}{d t} \left( \frac{ \zeta^2}{M^2}  \dot \psi \right)  + \left( k^2+ m^2_L \right) \psi =0 \, , \\
 m_L^2 =  \nabla_{\phi} \left[ \left(1 + \frac{4 \zeta^2}{M^2} \right) V_{\phi} \right]  \, . \label{m^2-L}
\eea
Simple inspection of this equation shows that indeed $| \ddot \chi | \ll M^2 | \chi|$ is satisfied. Additionally, from (\ref{v-N-chi}) notice that the vector (\ref{v-N-0}) is pointing almost entirely towards the direction $(1,0)$, which corresponds to the direction parallel to the motion of the background field. To deal with the previous equation we define
\be
e^\beta = 1+4 \zeta^2 /M^2 .
\ee
Then, we may write
\bea
 e^{\beta} \left( \ddot \psi   +  \dot \beta \dot \psi \right)  + \left(k^2 + m_L^2\right) \psi  =  0 \, , \\
 m_L^2 =  \nabla_{\phi} \left( e^\beta V_{\phi} \right) \, .
\eea
The mass term may be alternatively written as $m_L^2 =  \nabla_{\phi} \left( e^\beta V_{\phi} \right) = e^\beta \nabla_\phi V_\phi + e^\beta V_{\phi} \dot \beta / \dot \phi_0$. The previous equation of motion can be obtained from the action
\bea
 S = \frac{1}{2} \int dt d^3 x \left[  e^{\beta}  \dot \psi ^2 - \left( \nabla \psi \right)^2 - m_{L}^2 \psi^2  \right] \, .
\eea
By performing a field redefinition $\varphi \equiv e^{\beta/2} \psi$, we see that the previous action may be reexpressed as
\bea
 \label{low-energy-eff-action}
 S & = & \frac{1}{2} \int dt d^3 x \left[   \dot \varphi ^2 - e^{- \beta} ( \nabla \varphi)^2 - M_{L}^2 \, \varphi^2  \right]  , \quad \\
 M_{L}^2 & = & \nabla_\phi V_\phi + \frac{V_{\phi} \dot \beta }{ \dot \phi_0} - \frac{\ddot \beta}{2} - \frac{\dot \beta^2}{4}  .
\eea
Equation (\ref{low-energy-eff-action}) is one of our main results. It describes the precise way in which heavy physics manifests itself on low energy degrees of freedom when the full trajectory in scalar field space becomes non-geodesic. It was deduced by assuming $M^2 \gg k^2$ and remains valid for large values of $\beta$, which may be confirmed by comparing the effective theory with the results of Section \ref{section: constant radius of curvature}. For instance, in the particular case where $\nabla_\phi V_\phi = 0$ and the bending of the trajectory is such that $\dot \beta = 0$, one finds the frequency $\omega$ of the light mode to be given by $\omega = k e^{-\beta/2} = k M / \sqrt{M^2 + 4\zeta^2}$, which coincides with the previous result (\ref{omega-minus}). 

\section{Discussion}
\setcounter{equation}{0}
\label{conclusions}
In this note, we considered the structure of low energy description of scalar field
theories with a pronounced hie\-rar\-chy of mass scales. First, we set
up a framework for describing a light field moving along a multi-field
trajectory in field space. From this, we determined the background
equations of motion, around which we can study perturbations. Finally,
we deduced the effective theory describing light perturbations for the
case in which the background field is following a curved trajectory in
field space.

The main manifestation of the non-trivial mixing of the heavy and the
light directions is in the appearance of the coefficient $e^{-\beta}$
in front of the term $( \nabla \varphi)^2$ containing spatial
derivatives in the action
(\ref{low-energy-eff-action}). First, since $\beta \ge 0$,
the net effect of the bending of the background trajectory is to
reduce the energy per scalar field quantum. This is due to the fact
that during bending the light modes momentarily start exciting heavy
modes, therefore transferring energy to them. Second, since
$\beta = \ln ( 1 + 4\zeta^2 / M^2)$, there are two effects competing
against each other in this process. On one hand one has $\zeta = V_N /
\dot \phi_0$, or alternatively $ |\zeta| = \dot \phi / \kappa$, which
may be interpreted as the angular speed of the background field along
the curved trajectory. On the other hand, there is the mass of the
heavy mode $M$, which must be excited by the light modes during the
bending.

We stress that the magnitude of the effect is
parametrized entirely by the parameter $\beta$, which depends on the
velocity of the background trajectory and the induced curvature along
the scalar field trajectory through the field manifold, parametrized
by $\kappa$. Whether the kinetic terms are canonical or not is
immaterial as this is merely a question of the chosen basis in field
space.

In what follows we discuss two direct applications of our
results: inflation and the non-decoupling of light and heavy modes in
Supergravity.


\subsection{Inflation}
Understanding in detail how light and heavy modes remain coupled under
more general circumstances could be particularly significant for
cosmic
inflation~\cite{Guth:1980zm,Albrecht:1982wi,Linde:1981mu}. Indeed,
although current observations are consistent with the simplest model
of single field inflation, it is rather hard to conceive a realistic
model where the inflaton field alone is completely decoupled from UV
degrees of freedom. One way of addressing this issue is by studying
multi-field scenarios where many scalar fields have the chance to
participate in the inflationary dynamics~\cite{Starobinsky:1986fxa},
despite of different mass scales among the inflaton candidates. Hence,
there could exist certain phenomena related to inflation in which the
effects studied in this report can be relevant. This has also recently
been considered in Refs.~\cite{Tolley:2009fg, sera, Chen:2009zp}. 
A more detailed discussion can be found in
\cite{Achucarro:2010da}.

First, note that the equation of motion deduced from the
action (\ref{low-energy-eff-action}) is given by
\be \ddot \varphi + e^{-\beta} k^2 \varphi + M^2_L \, \varphi = 0 \, .
 \label{effective-eq-2}
\ee
For definiteness, let us focus on phenomena characterized by $\left|
  \dot \beta \right| \ll k$ and consider the case in which the
potential is flat enough so that $M^2_L \ll k^2$ is satisfied. Then,
the time variation of $\beta$ along the trajectory is small enough to
allow us to write the mode solution as
\be
 \varphi (k,t) = \frac{e^{\beta/4}}{\sqrt{k}} \exp \left[ i e^{-\beta/2} k \, t \right] \, ,
 \label{sol-beta}
\ee
where the factor $e^{\beta/4}/\sqrt{k}$ is necessary in order to
satisfy the commutation relation $[ \varphi, \dot \varphi ] = i$. This
factor coincides with the one found in (\ref{expansion-v-1}) for the
amplitude of light modes in the case where $\beta$ is a constant. To
continue, from (\ref{sol-beta}) we can see that in the vacuum, the two
point correlation function of the perturbation $\varphi ({\bf x}, t)$,
has the form $\langle \varphi ({\bf x}, t) \varphi ({\bf y},t) \rangle
\propto e^{\beta(t)/2}$. One direct consequence of this result is for
inflation, where the amplitudes of scalar fluctuations freeze after
crossing the horizon, i.e. when the physical wavelength $k^{-1}$
satisfies the condition $e^{- \beta/2} k = H$. More precisely, if we
generalize (\ref{effective-eq-2}) to include gravity, we would
conclude that the speed of sound of adiabatic perturbations is given
by
\be
 c_s^2 = e^{- \beta} \, .
\ee
Such an effect is known to produce sizable levels of non-Gaussianities~\cite{Alishahiha:2004eh}. Additionally it modifies the power spectrum as
\be
 P(k) \simeq e^{\beta(k) / 2} P_*(k) \, ,  \label{power-beta-2}
\ee
where $P_*(k)$ is the conventional power spectrum $P_*(k) \propto
k^{n_s -1}$ deduced in single field slow-roll inflation, and
$\beta(k)$ is the value of $\beta(t)$ at the time $t$ when the mode
$k$ crosses the horizon~\footnote{In the constant curvature case
  $\kappa=\textrm{const.}$, $\beta(\kappa)$ is constant as well and we
  find an overall modulation of the power spectrum compatible with
  Ref.~\cite{Chen:2009zp}.}. Certainly the more interesting case
corresponds to a varying $\beta$, and since $\beta$ can be large, such
an effect may be sizable and observable in the near future. In many
scalar field theories, such as supergravity, the masses of heavy
degrees of freedom during inflation are typically of order $M \sim H$,
leading to the relation
\be
e^\beta \sim 1+ 4 \epsilon  \frac{M_{\rm Pl}^{2}}{ \kappa^2} \, .
\ee
If the bending is such that $\kappa \ll M_{\rm Pl}$ (a feasible
situation) one then could obtains effects as large as $\beta \sim
1$. In the case where a turn of the trajectory happens during a few
$e$-folds, one then should be able to observe features in the power
spectrum of $\mathcal{O}(1)$, particularly by modifying the running of
the spectral index as $d n_s / d \ln k$, which otherwise would be of
$\mathcal{O}(\epsilon^2)$. A detailed computation of this effect can be found in ~\cite{Achucarro:2010da}, where the
interaction between light an heavy modes during horizon crossing is
examined more closely (see also \cite{sera}).


\subsection{Decoupling of light and heavy modes in supergravity}
Our results can be also used to assess when a low energy effective theory, deduced from a multi-scalar field theory containing both heavy directions and light directions, is accurate enough. As discussed in full detail in Appendix \ref{appendix-zero-order}, whenever the background fields are evolving (as in inflation) the only way of having a vanishing $\beta$-parameter is for a trajectory to correspond to a geodesic in the full scalar field manifold $\mathcal M$. It is clear that the only way of achieving this is by having some property relating the shape of the potential $V(\phi)$ with the geometry of $\mathcal{M}$. In the particular case of supergravity such a property is known to exist, and therefore one should expect supergravity theories rendering low energy effective theories for which $\beta$ vanishes exactly. To be more precise, in $N=1$ supergravity the scalar field potential is given by
\be
 V = e^{G} \left(\gamma^{a \bar b} G_a G_{\bar b} - 3\right) \ ,
\ee
where $\gamma^{a \bar b}$ is the inverse of the K\"ahler metric
$\gamma_{a \bar b} = \partial_{a} \partial_{\bar b} G$ deduced out of
the generalized K\"ahler potential $G$, which is a real function of the
chiral scalar fields $\phi^a$, $G_a = \partial_{a} G$ and $G_{\bar b}
= \partial_{\bar b} G$. Now, consider a supergravity theory in which a
set of massive chiral fields $\phi^H$ satisfy the condition
\be
 G_H = 0 \ ,
\ee
along a given hypersurface $\mathcal S$ in $\mathcal{M}$ parametrized
\emph{only} by the light fields $L$. This means that a surface $S$ is
defined by $f(H,\bar{H})=0$ rather that by a function
$f(H,\bar{H},L,\bar{L})=0$~\cite{deAlwis:2005tf} (see
also~\cite{Achucarro:2008sy}). Then it is possible to verify that the
scalar fluctuations $\phi^L$ parallel to $\mathcal S$ are decoupled
from the fields $\phi^H$, rendering $\beta = 0$. To appreciate this,
observe first that at any point on the surface $\mathcal S$ the
K\"ahler metric $\gamma_{a \bar b}$ is diagonal between the two
sectors. Indeed, since $G_H = 0$ holds at any point in the surface,
then it must be independent of arbitrary displacements $\delta \phi^L$
along $\mathcal S$. This implies that
\be
 \partial_{\bar L} G_H = \gamma_{H \bar L}  = 0  \qquad {\rm on} \,\, \mathcal S.
\ee
This condition automatically ensures that in the absence of a scalar
field potential, the trajectory along $\mathcal S$ will be a
geodesic. It remains then to verify that the potential does not imply
quadratic couplings between both sectors, and therefore
leaves these geodesic trajectories unmodified. Consider the first and
second derivatives of the potential
\begin{widetext}
  \bea
  \label{der-1}
  \nabla_a V &=&  e^{G} \Big(G_a + G^b \nabla_a G_b \Big) + V G_a \, , \\
  \label{der-2}
  \nabla_a \nabla_{\bar b} V &=&  e^{G} \Big(\gamma_{a \bar b} + \! \nabla_a G_c \nabla_{\bar b} G^c \! - \mathcal{R}_{a \bar b c \bar d} G^c G^{\bar d}   \Big) \! +   G_a V_{\bar b}   +  G_{\bar b} V_a  +   ( \gamma_{a \bar b} - G_a G_{\bar b} )  V \, ,  \quad  \\
  \label{der-3}
  \nabla_a \nabla_b V &=& e^{G} \Big( 2 \nabla_{a} G_{b} + G^c
  \nabla_a \nabla_b G_c \Big) + G_a V_{b} + G_{b} V_a + (\nabla_{a}
  G_{b} - G_a G_{b} ) V \, .  \eea
\end{widetext}
Since $\gamma_{H \bar L} = 0$ and $G_H=0$, from (\ref{der-1}) one
immediately obtains $V_H = 0$ in $\mathcal S$. It is not difficult to
notice that also $\nabla_L V_H = \nabla_{\bar L} V_H = \nabla_{L}
V_{\bar H} = \nabla_{\bar L} V_{\bar H} = 0$ which also hinge on
$\gamma_{H \bar L} = 0$ and $G_H=0$. All of these results put together
imply that the heavy sector will not affect the light sector as long
as the background trajectory remains on the geodesically
generated surface $\mathcal S$. Conversely, deviations from the
condition $G_{H} = 0$, or a surface $S$ with $H \neq \mathrm{const.}$,
will produce interactions leading to the appearance of the coupling
$\beta$ studied in the present work. A particularly interesting
example is Ref.~\cite{Gallego:2008qi}, where $O(\epsilon)$ couplings
between heavy and light fields in the superpotential result in
suppressed, $O(\epsilon^{2})$ terms in the effective action for the
light fields. This result was obtained by expanding about a particular
$H=\mathrm{const.}$ configuration which, for constant light background
fields, only deviates at $O(\epsilon)$ from the true solution to the
equations of motion. But along an arbitrary background $L(t)$ the
deviation will exceed $O(\epsilon)$ for displacements $\Delta L /
\kappa > \epsilon$ (due to the $\Gamma^{H}_{LL} \dot{L}^{2}$ term in
the $H$ equation of motion), and the corrections to the effective
action discussed in this paper become dominant.

\section*{Acknowledgements}
We would like to thank Koenraad Schalm and Ted van der Aalst for discussion and comments. This work is supported by the NWO under the VIDI and VICI programs (AA,JG,SH,GAP), by Conicyt under the Fondecyt Initiation Research Project 11090279 (GAP), by funds from CEFIPRA/IFCPAR (SP) and by project CPAN CSD2007-D004 (AA). SP wishes to thank the theory group at Leiden University and the ISCAP at Columbia University for hospitalities during preparation of the manuscript.

\begin{appendix}
\renewcommand{\theequation}{\Alph{section}.\arabic{equation}}
\setcounter{section}{0}
\setcounter{equation}{0}

\section{Geometric identities}
\label{sec:app-geom}
\setcounter{equation}{0}
In this appendix, we derive some additional useful geometric identities. As before, we study a system characterized by a manifold $\mathcal{M}$ of dimension $n_{\rm tot}$ with metric $\gamma_{a b}$. The Christoffel connections are given by
\be
 \Gamma^{a}_{b c} = \frac{1}{2} \gamma^{a d} \left( \partial_b \gamma_{d c} + \partial_c \gamma_{b d} - \partial_d \gamma_{b c} \right) \, ,
\ee
with $\partial_{a}$ the partial derivatives with respect to $\phi^a$. From this, we define the covariant derivatives  $\nabla_a X_b \equiv \partial_a X_b - \Gamma_{a b}^c X_c$. Then, the Riemann tensor is
\be
 \mathcal{R}^a_{\,\,\, b c d} \equiv \partial_c \Gamma^a_{b d} - \partial_d \Gamma^a_{b c} + \Gamma^a_{c e} \Gamma^e_{d b} -   \Gamma^A_{d e} \Gamma^e_{c b} \, ,
 \label{Riemann-def-app}
\ee
which can be used to define $\mathcal{R}_{ab}=\mathcal{R}^{c}_{\phantom{c}acb}$ and $\mathcal{R}=\gamma^{ab}\mathcal{R}_{ab}$.

If $\dot{\phi}\neq0$ we define vectors parallel and orthogonal to the field trajectory $\phi_{0}$ as before,
\bea
 T^a & \equiv & \frac{\dot \phi_0^a}{\dot \phi_0} \, ,
 \\
 N^a & \equiv & s_N(t) \left(  \gamma_{bc} \frac{D T^b}{d t} \frac{DT^c}{dt} \right)^{-1/2 }\frac{D T^a}{dt} \, , 
 \label{def-T-N-app}
\eea
where $s_N(t)$ is defined in the way explained in Section~\ref{sec:background}.
With the help of the equation of motion (\ref{scal-eq-1}) we can then show that
\bea
 \frac{D T^a}{dt} = - \frac{\ddot \phi_0}{\dot \phi_0} T^a - \frac{1}{\dot \phi_0}  V^a \, ,
 \label{alt-eq-1-app}
\eea
which, projected along  $T^a$ and $N^a$,  leads to two independent equations of motion
\bea
 \label{eq-mot-sigma-app}
 \ddot \phi_0 + V_{\phi} & = & 0 \, , \\
 \frac{D T^a}{dt} & = & - \frac{V_N}{\dot \phi_0} N^a \, .
 \label{eq: Dtdt-Vs-app}
\eea

From this, we can characterize the time variation of $N^a$. Taking a total time derivative to (\ref{scal-eq-1}) we obtain
\be
 \frac{1}{\dot \phi_0} \frac{D^2 \dot \phi^a_0}{dt^2} = -  \nabla_{\phi} V^a \, ,
\ee
where $\nabla_{\phi} \equiv T^a \nabla_{a}$ is the covariant derivative along the trajectory. Using the definition of $T^a$, the previous expression can be used to compute
\be
 \frac{D^2 T^a}{dt^2} = \left( \nabla_{\phi} V_{\phi} \right) T^a - \nabla_{\phi} V^{a} - \frac{V_{\phi} V_N}{\dot \phi_0^2} N^a \, .
\ee
Then, by further recalling the definition (\ref{def-T-N-app}) of $N^a$ and inserting it on the left hand side of the previous expression, one arrives to
\be
 \frac{D  N^a}{dt} = \frac{V_N}{\dot \phi_0} T^a + \frac{\dot \phi_0}{V_N}  P^{a b} \nabla_{\phi} V_b \, , \label{DN-P}
\ee
where we have defined the projector tensor $P^{a b} \equiv \gamma^{a b} - T^a T^b - N^a N^b$ along the space orthogonal to the subspace spanned by the unit vectors $T^a$ and $N^a$. Observe that in the particular case of two-field models $n_{\rm tot} = 2$, one has $\gamma_{a b} = T_a T_b + N_a N_b$ and therefore  $P^{a b}= 0$.  The radius of curvature $\kappa$ of the trajectory followed by the vacuum state in the scalar manifold $\mathcal{M}$ is defined as
\be
 \frac{1}{\dot \phi_0} \frac{D T^a}{d t} = \frac{D T^a}{d \phi} = s_N(t) \frac{N^a}{\kappa}  \, ,
\ee
Finally, using~(\ref{eq: Dtdt-Vs-app}), we can deduce (eq.~\ref{kappa-def})
\be
\frac{1}{ \kappa} = \frac{|V_N|}{\dot \phi_0^2}\, .
\ee

\section{Initial conditions for perturbations}
\label{sec:app-init-cond}
\setcounter{equation}{0}
In this appendix we study the conditions that the $n_{\rm tot}$ mode solutions $v_{ \alpha }^I (k,t)$ defined in (\ref{v-fourier}) must satisfy to be compatible with the commutation relations (\ref{commutation-1}). We find that the conditions on the $v_{ \alpha }^I (k,t)$ are
\bea
 \label{cond-v-1}
 \sum_{\alpha}  \left[ v_{\alpha}^I  \frac{ \mathcal D v_{\alpha}^{J *}}{dt}   -  \frac{ \mathcal D v_{\alpha}^{J}}{dt}  v_{\alpha}^{I *}  \right] &=& i \delta^{I J} \, , \\
 \label{cond-v-2}
 \sum_{\alpha}  \left[ v_{\alpha}^I  v_{\alpha}^{J *} -  v_{\alpha}^J  v_{\alpha}^{I *}  \right] &=& 0~, \\
 \label{cond-v-3}
 \sum_{\alpha}  \left[ \frac{ \mathcal D v_{\alpha}^{I}}{dt}  \frac{ \mathcal D v_{\alpha}^{J *}}{dt}   - \frac{ \mathcal D v_{\alpha}^{J}}{dt}  \frac{ \mathcal D v_{\alpha}^{I *}}{dt}  \right] &=& 0 \, .
\eea
To show that these relations can be imposed at any given time $t$ we proceed as follows. First, let us define the following matrices:
\bea
 A^{I J} &=& i \sum_{\alpha} \left[ v_{\alpha}^I  v_{\alpha}^{J *} -  v_{\alpha}^J  v_{\alpha}^{I *}  \right] \, , \\
 B^{I J} &=& i \sum_{\alpha} \left[ \frac{ \mathcal D v_{\alpha}^{I}}{dt}  \frac{ \mathcal D v_{\alpha}^{J *}}{dt} - \frac{ \mathcal D v_{\alpha}^{J}}{dt}  \frac{ \mathcal D v_{\alpha}^{I *}}{dt}  \right] \, , \\
 E^{I J} &=& i \sum_{\alpha}  \left[ v_{\alpha}^I  \frac{ \mathcal D v_{\alpha}^{J *}}{dt}   -  \frac{ \mathcal D v_{\alpha}^{J}}{dt}  v_{\alpha}^{I *}  \right] \, .
\eea
It is possible to show that they satisfy the following properties:
\bea
 A^{IJ} & = & A^{IJ *} = - A^{JI} \, , \\
 B^{IJ} & = & B^{IJ*} = - B^{JI} \, , \\
 E^{IJ} & = & E^{I J*} \, .
\eea
That is, all of them are real, and $A^{IJ}$ and $B^{IJ}$ are antisymmetric. Due to these properties it is possible to appreciate that $A^{IJ}$ and $B^{IJ}$ consist of $n_{\rm tot} (n_{\rm tot}-1)/2$ independent real components each. On the other hand, $E^{IJ}$ consists of $n_{\rm tot}^2$ independent real components. Therefore, In order to fix the values of all of them we need to specify $2 n_{\rm tot}^2 - n_{\rm tot}$ independent quantities. They also satisfy the following equations of motion:
\bea
 \label{tensor-1}
 \frac{ \mathcal D}{dt} A^{IJ} &=& E^{IJ} - E^{JI} \, , \\
 \label{tensor-2}
 \frac{ \mathcal D}{dt}  B^{IJ} &=& C^{I}_{\phantom{I} K}E^{KJ} -  C^{J}_{\phantom{J} K}E^{K I} \, , \\
 \label{tensor-3}
 \frac{\mathcal D}{dt} E^{IJ} &=& B^{IJ} + A^{IK} \left( k^2 \delta^J{}_K + C_{K}^{\phantom{K} J} \right)~.
\eea
Taking the trace to the last equation, we obtain that the trace $E \equiv E^{I}_{\phantom{I} I}$ satisfies
\be
 \frac{d E}{d t} = 0 \, ,
\ee
and therefore $E$ is a constant of motion of the system. Furthermore, observe that the configuration
\bea
 \label{quantum-conditions1}
  E^{IJ} & = & \frac{E \delta^{IJ}}{n_{\rm tot}} \, , 
  \\
 \label{quantum-conditions2}
  A^{IJ} & = & B^{IJ} = 0 \, ,
\eea
for which conditions (\ref{cond-v-1}) to (\ref{cond-v-3}) are satisfied corresponds to a fixed point of the set of equations (\ref{tensor-1}) to (\ref{tensor-3}). That is, they automatically satisfy $\mathcal DA^{IJ}/dt = \mathcal D B^{IJ}/dt = \mathcal D E^{IJ}/dt = 0$. Therefore, it remains to verify whether there exist sufficient independent degrees of freedom in order to satisfy the initial conditions $E^{IJ} = E \delta^{IJ} / n_{\rm tot}$ and $A^{IJ} = B^{IJ} = 0$ at a given initial time $t_{0}$. As a matter of fact, we have exactly the right number of degrees of freedom. As we have already noticed there exists $n_{\rm tot}$ independent solutions $v_{\alpha}^{I} (t , k)$ to the equations of motion. To fix each solution $v_{\alpha}^{I} (t , k)$ we therefore need to specify $2 n_{\rm to t}^2$ independent quantities, corresponding to the addition of $n_{\rm tot}^2$ components $v_{\alpha}^{I} (t_0)$ and $n_{\rm tot}^2$  momenta $\mathcal{D} v_{\alpha }^{i} (t_0)/dt$. However
  we must notice that the overall phase of each solution $v_{\alpha}^{i} (t , k)$ plays no roll in satisfying the initial values for $A^{IJ}$, $B^{IJ}$ and $E^{IJ}$. We therefore have precisely $2 n_{\rm tot}^2 - n_{\rm tot}$ free parameters to set $E^{IJ} = E \delta^{IJ} / n_{\rm tot}$ and $A^{IJ} = B^{IJ} = 0$. Of course, the value of the trace $E$ is part of this freedom, and we are free to set it in such a way that  $E / n_{\rm tot} = 1$.

\section{Zeroth-order theory of the background fields}
\label{appendix-zero-order}
\setcounter{equation}{0}
In this appendix we study in detail the dynamics offered by the tree level potential $V(\phi) = V_* (\phi)$ discussed in Section \ref{Hierarchies}. We shall focus only on potentials $V$ for which the Hessian $V_{a b}$ is positive definite. Let us for a moment independently consider solutions to the equation
\be
 V^a = 0 \, .
\ee
In general, these will correspond to a set of fields parametrizing a surface $\mathcal{S}$ in $\mathcal{M}$. The fields lying on this surface correspond to exactly flat directions of the potential $V$. Let us express this surface by means of the following parametrization
\be
 \phi_*^a = \phi_*^a(\chi^{\alpha}) \, ,
 \label{min-sol}
\ee
where $\alpha = 1, \cdots n_{\mathcal{S}}$, with $n_{\mathcal S}$ the number of flat directions of the potential. Then
\be
 V_a \left[\phi_*(\chi)\right] = 0
\ee
for any $\chi$. Clearly, $n_{\mathcal S}$ is the dimension of the surface. We may now define the induced metric on the surface by making use of the pullbacks $X^a_{\,\,\, \alpha} \equiv \partial_{\alpha} \phi_*^a$ in the following way
\be
 g_{\alpha \beta} = X^a_{\,\,\, \alpha} X^b_{\,\,\, \beta} \gamma_{a b } \, .
\ee
Let us for a moment disregard the degrees of freedom perpendicular to this surface and consider only those lying on $\mathcal{S}$. This corresponds to truncate the theory by considering only the fields $\chi^\alpha$. The theory for such fields would be deduced from the action
\be
 S = - \frac{1}{2}  \int \!  d^4 x \, g_{\alpha \beta} \partial_{\mu} \chi^\alpha  \partial^{\mu} \chi^\beta \, ,
\ee
and the equations of motion would be given by
\be
 \frac{D}{dt} \dot \chi^\alpha =  \frac{d^2 \chi^{\alpha} }{d t^2} + \hat \Gamma^{\alpha}_{\beta \gamma} \frac{d \chi^\beta}{d t}  \frac{d \chi^\gamma}{d t} = 0 \, ,
 \label{eq-for-light-trunc}
\ee
where
\be
 \hat \Gamma^{\alpha}_{\beta \gamma} = \frac{1}{2} g^{\alpha \delta} \left( \partial_\beta g_{\delta \gamma} + \partial_{\gamma} g_{\beta \delta} - \partial_\delta g_{\beta \gamma} \right)
\ee
is the connection deduced out of the induced metric $g_{\alpha \beta}$. The relation between $\hat \Gamma^{\alpha}_{\beta \gamma}$ and $\Gamma^{a}_{b c}$ is given by
\be
 \hat \Gamma^{\alpha}_{\beta \gamma} = X_a^{\,\,\, \alpha} \left( X^b_{\,\,\, \beta} X^c_{\,\,\, \gamma} \Gamma_{b c}^a + X^{a}_{\,\,\, \beta \gamma} \right) \, ,
\ee
where $X^{a}_{\,\,\, \beta \gamma} \equiv \partial_\gamma X^a_{\,\,\, \beta}$. It is convenient here to define $M^a_{\beta \gamma} \equiv X^b_{\,\,\, \beta} X^c_{\,\,\, \gamma} \Gamma_{b c}^a + X^{a}_{\,\,\, \beta \gamma} $. Then, one has $\hat \Gamma^{\alpha}_{\beta \gamma} = X_a^{\,\,\, \alpha} M^a_{\beta \gamma}$. Let us review under what conditions the previous truncation is consistent. For this, let us recall how much a solution to (\ref{eq-for-light-trunc}) deviates from the equation of motion of the full theory given by (\ref{scal-eq-1}). By differentiating with respect to time the solution~(\ref{min-sol}) with $\chi^\alpha$ satisfying (\ref{eq-for-light-trunc})  we find
\bea
 \frac{D}{dt} \dot  \phi_*^a(\chi) &=& X^a_{\,\,\, \alpha}  \ddot \chi^{\alpha} +  M^a_{\alpha \beta} \dot \chi^{\alpha} \dot \chi^{\beta} , 
\eea
or alternatively:
\bea
\quad \frac{D}{dt} \dot  \phi_*^a(\chi)   &=&  \left( M^a_{\alpha \beta} -  X^a_{\,\,\, \gamma} X_b^{\,\,\, \gamma} M^b_{\alpha \beta} \right) \dot \chi^{\alpha} \dot \chi^{\beta} \, . \qquad
 \label{non-geo}
\eea
It is useful to define $Q^a_{\alpha \beta} \equiv P^a_b M^b_{\alpha \beta} $, where $P^a{}_b \equiv \delta^a{}_b - X^a_{\,\,\, \gamma} X_b^{\,\,\, \gamma} $ is the projector along the space perpendicular to the surface.  $Q^a_{\alpha \beta}$ transforms as a tensor:
\be
 Q^a_{\alpha \beta} = \partial_\alpha X^{a}_{\,\,\, \beta} + \Gamma^a_{b \alpha} X^{b}_\beta - \hat \Gamma^{\gamma}_{\alpha \beta} X^{a}_{\,\,\, \gamma} = D_{\alpha} X^{a}_{\,\,\, \beta} \, ,
\ee
where $\Gamma^a_{b \alpha} \equiv \Gamma^a_{b c} X^{c}_{\,\,\, \alpha}$. The previous notation is consistent as $X^{a}_{\,\,\, \alpha}$ transforms homogeneously under reparametrizations of $\phi$ and $\chi$. Thus, finally we are left with
\bea
 \frac{D}{dt} \dot  \phi_*^a(\chi)   =  Q^a_{\alpha \beta}  \dot \chi^{\alpha} \dot \chi^{\beta} \, .
 \label{non-geo-2}
\eea
Therefore, since $V^a(\phi_*) = 0$ by definition, if $Q^a_{\alpha \beta}  \dot \chi^{\alpha} \dot \chi^{\beta}$ is non-vanishing along the trajectory followed by $\chi^{\alpha}$, then $\phi_{*}^a$ does not satisfy the equations of motion for $\phi^a$ in the full theory. In fact, we are interested in an arbitrary solution $\chi^{\alpha} = \chi^{\alpha}(t)$ of (\ref{eq-for-light-trunc}), thus in general, either $\dot \chi^\alpha = 0$ or $Q^a_{\alpha \beta} = 0$. The first case corresponds to a stationary solution, where the background is not evolving. The second case $Q^a_{\alpha \beta} = 0$ is more interesting, as it corresponds to the case in which $\mathcal{S}$ is geodesically generated. To appreciate this, notice first that if $Q^a_{\alpha \beta} = 0$ then $\phi_*^a = \phi_*^a(t)$ satisfies the equation of a geodesic. In second place, it is possible to deduce the following identity
\bea
 \mathcal{R}^a_{\phantom{a} \alpha \beta \gamma} &\equiv & P^a{}_b X^c_{\,\,\, \alpha} X^d_{\,\,\, \beta} X^e_{\,\,\, \gamma} \mathcal{R}^{B}_{\phantom{B} c d e}  \nonumber \\
 &=& P^a{}_b \left(D_\beta Q^b_{\gamma \alpha} - D_\gamma Q^b_{\beta \alpha}\right) \, .
\eea
Thus, if  $Q^a_{\alpha \beta} = 0$ then arbitrary vectors, which are tangent to $\mathcal{S}$, will not generate a component normal to $\mathcal{S}$ after being transported around an arbitrary loop in $\mathcal{S}$. Finally, one also has the general relation
\bea
 \hat \mathcal{R}^{\alpha}_{\,\,\, \beta \gamma \delta} &=& X_{a}^{\,\,\, \alpha}  X^b_{\,\,\, \beta} X^c_{\,\,\, \gamma} X^d_{\,\,\, \delta} \mathcal{R}^a_{\phantom{a} b c d}  \nn\\ 
 && + \left( Q^{a}_{\beta \delta} \gamma_{a b} Q^{b}_{\sigma \gamma} g^{\sigma \alpha} -  Q^{a}_{\beta \gamma} \gamma_{a b} Q^{b}_{\sigma \delta} g^{\sigma \alpha} \right)  , \qquad
\eea
meaning that if $Q^a_{\alpha \beta} = 0$ one has that the Riemann tensor $\hat \mathcal{R}^{\alpha}_{\,\,\, \beta \gamma \delta}$ characterizing $\mathcal{S}$ coincides with the induced Riemann tensor $X_{a}^{\,\,\, \alpha}  X^b_{\,\,\, \beta} X^c_{\,\,\, \gamma} X^d_{\,\,\, \delta} \mathcal{R}^a_{\phantom{a} b c d}$ to the surface.

\begin{figure}
\includegraphics[width=0.4\textwidth]{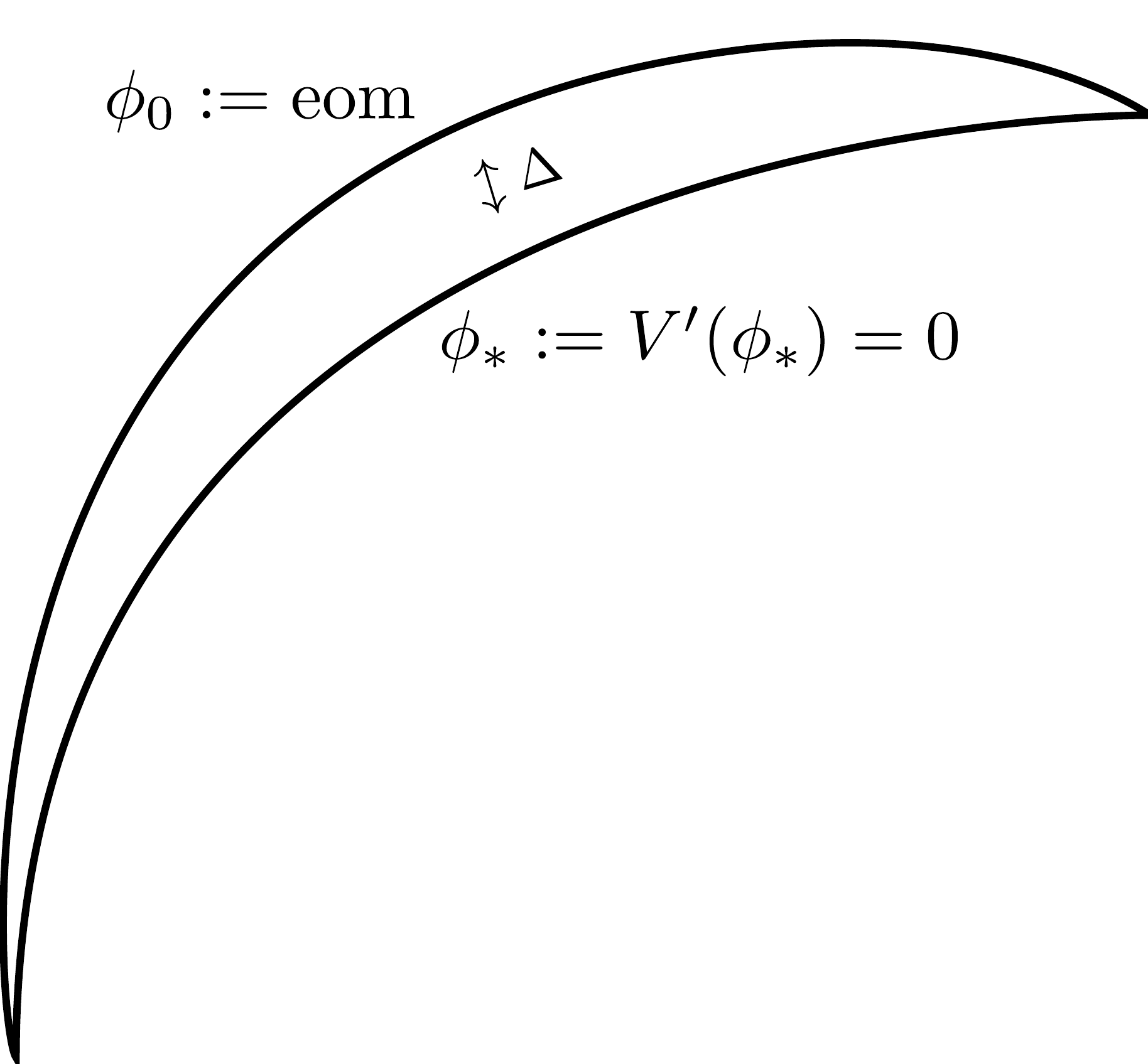}
\caption{The difference between $\phi_{*}$ and $\phi_{0}$.}
\label{fig:ex_traj}
\end{figure}

It is rather clear that whenever the surface $\mathcal{S}$ is not geodesically generated, the solution $\phi^A = \phi^A(\chi)$ is not a solution of the full set of equations of motion. Let us now ask under what circumstances this might be a good approximation. For this, consider the following notation for the full solution
\be
 \phi^a = \phi^a_* + \Delta^a \, ,
\ee
where $\Delta^a$ has the purpose of parametrizing the displacement of the full solution from $\phi^a_*$ defining the surface $\mathcal{S}$ (See figure \ref{fig:ex_traj}). 
To deduce the equation of motion for $\Delta^a$ notice that
\bea
 \frac{D \dot \phi^a}{d t} &=&  \ddot \phi^a + \Gamma^{a}_{b c}(\phi) \dot \phi^b \dot \phi^c 
 \nonumber\\
 &=& \ddot \phi^a_* + \ddot \Delta^a + \Gamma^{a}_{b c} (\phi_* + \Delta) \left(\dot \phi_* + \dot \Delta\right)^b \left(\dot \phi_* + \dot \Delta\right)^c
 \nonumber\\
 &=&  \frac{D \dot \phi_*^a}{d t} + \ddot \Delta^a  + \Gamma^{a}_{b c} (\phi_*)  \dot \Delta^b   \dot \phi_*^c + \Gamma^{a}_{b c} (\phi_*)   \dot \phi_*^b \dot \Delta^c \nn\\ 
 && + \partial_d \Gamma^{a}_{b c} (\phi_*) \dot \phi_*^b \dot \phi_*^c \Delta^d \, ,
\eea
where we kept terms up to order $\Delta$. On the other hand, we have the relation
\bea
 \frac{D^2 \Delta^a}{dt^2} &=& \left[\dot \Delta^a + \Gamma^{a}_{b c} (\phi_*) \Delta^b \dot \phi_*^c \right]\dot{} \nn\\
 && + \Gamma_{b c}^A(\phi_*) \left[\dot \Delta^b + \Gamma^{b}_{de} (\phi_*) \Delta^d \dot \phi_*^e\right] \dot \phi_*^c \, . \quad
\eea
Putting these two expressions together we find the equation of motion for $\Delta^a$ to be given by
\be
 \frac{D^2 \Delta^a}{dt} + Q^a_{\alpha \beta} \dot \chi^\alpha \dot \chi^\beta + C^a_{\phantom{a} b} (\phi_*) \Delta^b = 0 \, ,
 \label{eq-Delta}
\ee
where we are neglecting terms of higher order in $\Delta$. In the previous expression we have defined
\be
 C^a_{\phantom{a} b} (\phi_*) \equiv V^a_{\phantom{a} b} (\phi_*) - \mathcal{R}^a_{\phantom{a} c d b} (\phi_*) \dot \phi_*^c \dot \phi_*^d \, ,
\ee
where $V^a_{\phantom{a} b} (\phi_*) \equiv \gamma^{a c}(\phi_*) \nabla_{c} V_{b} (\phi_*) $. In deriving this expression we have  assumed that $Q^a_{\alpha \beta} \dot \chi^\alpha \dot \chi^\beta$ is of $\mathcal{O}(\Delta)$. This is correct since we need to demand $\Delta = 0$ for the particular case $Q^a_{\alpha \beta} \dot \chi^\alpha \dot \chi^\beta = 0$. That is to say, we are strictly interested in the inhomogeneous solution of the previous equation (we shall later address the issue regarding perturbations). Notice that the effective mass $C^{a}_{\phantom{a}b}$ contains a contribution from the Riemann tensor. However, the direction given by $\dot \phi_{*}^a$ continues to be a flat direction since $\mathcal{R}^a_{\phantom{a} c d b} (\phi_*) \dot \phi_*^c \dot \phi_*^d \dot \phi_*^b = 0$. In other words,
\be
 C^a_{\phantom{a} b} (\phi_*) \dot \phi_*^b = 0 \, .
\ee
Additionally, notice that $C_{a b}$ is symmetric. To proceed, let us define a few more quantities. First, the tangent vector to the trajectory defined by $\phi_{*}(t)$ on the surface is given by
\be
 T_*^{a} = \frac{\dot \phi_*^a}{\dot \phi_*} \, ,
\ee
where $ \dot \phi_*^2 = \gamma_{a b} \dot \phi_*^a \dot \phi_*^b $. In fact, notice that
\bea
 T_*^{a} & = & X^a_{\,\,\, \alpha} T_*^\alpha \, , 
 \\
 T_*^\alpha & = & \frac{\dot \chi^\alpha}{\dot \phi_*} \, ,
 \\
 \dot \phi_*^2 & = & g_{\alpha \beta} \dot \chi^\alpha  \dot \chi^\beta \, .
\eea
It is a simple matter to show that
\bea
 \frac{D T_*^a}{d t} & = & \dot \phi_* Q^a_{\alpha \beta} T_*^\alpha T_*^\beta \, ,
 \\
 \ddot \phi_* & = & 0 \, .
\eea
It follows that $N_*^a \propto Q^a_{\alpha \beta} T_*^\alpha T_*^\beta $. It should be clear that $T_*^b V^a_{\phantom{a} b} (\phi_*) = 0$, as $T_*^a$ is by definition along the flat directions of the potential. It is useful to consider the definition of the radius of curvature $\kappa_{*}$ parametrizing the deviation of the trajectory in $\mathcal{S}$ with respect to geodesics in $\mathcal{M}$. The radius of curvature $\kappa_{*}$ comes defined as
\be
 \frac{D T_*^a}{d \phi_*} =  s_N(t) \frac{N_*^a}{\kappa_*}  \, ,
\ee
and therefore one has
\be
 \frac{1}{\kappa_*} =   | N_{*a} Q^a_{\alpha \beta} T_*^\alpha T_*^\beta | =  \sqrt{\gamma_{a b } Q^a_{\alpha \beta} T_*^\alpha T_*^\beta Q^b_{\gamma \delta} T_*^\gamma T_*^\delta} \, .
\ee
Notice that this quantity depends only on geometrical objects, as it should. Coming back to (\ref{eq-Delta}), we may now write
\be
 \frac{D^2 \Delta^a}{dt^2} - s_N(t) \dot \phi_*^2 N_*^a \kappa_*^{-1}  + C^a_{\phantom{a} b} (\phi_*) \Delta^b = 0~.
\ee
At this point one may argue that there are no good reasons to consider $\kappa_{*}^{-1}$ to be a small parameter. In fact, typically, for theories incorporating modular fields, $\kappa$ may be of $\mathcal{O}(1)$ in Planck unit or even smaller. Since $\dot \phi_*$ is constant, it is convenient to parametrize the trajectory with $\phi_*$. We can in fact write
\bea
 \frac{D \Delta^a}{d t} & = & \dot \phi_* \frac{D \Delta^a}{d \phi_*} \, ,
 \\
 \frac{D^2 \Delta^a}{d t^2} & = & \dot \phi_*^2 \frac{D^2 \Delta^a}{d \phi_*^2} \, .
\eea
We can therefore reexpress the equation of motion for $\Delta^a$ in terms of the proper parameter $\phi_*$ along the curve
\be
 \frac{D^2 \Delta^a}{d \phi_*^2}  + \frac{1}{\dot \phi_*^2}C^a_{\phantom{a} b} (\phi_*) \Delta^b =   s_N (\phi_*) N_*^a \kappa_*^{-1} \, ,
 \label{eq-Delta-2}
\ee
where now $ s_N(\phi_*)$ denotes the sign of $N_*^a $ relative to $ D T_*^a / d \phi_* $ but expressed as a function of $\phi_*$ instead of time.
To gain experience with this equation, consider the following situation. Suppose we have a trajectory in field space characterized by a constant curvature $\kappa$ and such that $C^a_{\phantom{a}b} N^b = M^2 \, N^a$ with $M^2>0$ a constant. That is, $N^a$ is an eigenvector of $C^a_{\phantom{a} b}$. Under such conditions, using the results of Section~\ref{sec:background} we find that
\be
 \frac{D^2 N_*^a}{d \phi_*^2} =  - \frac{N_*^a}{\kappa_*^2}  \, .
\ee
Then, we can see that $\Delta^a = \Delta \, N^a$ with $\Delta$ constant is a solution of the equation, with
\be
 \Delta = \frac{\dot \phi_*^2}{\kappa_{*}} \left( M^2 - \frac{\dot \phi_*^2}{\kappa_*^2} \right)^{-1} \, .
\ee
If $M^2 \gg \dot \phi_*^2 / \kappa_{*}^{2}$, which corresponds to the case in which the energy scale of the low energy dynamics is much smaller than the energy scale associated to the heavy fields, we simply have
\be
 \Delta \simeq \frac{\dot \phi_*^2}{M^2  \kappa_*} \, ,\label{eq-Delta-4}
\ee
This is the typical deviation from the true minimum of the potential if the surface of this minimum is not geodesic. To be more general, let us focus on a class of background trajectories in which
\be
 \frac{D \Delta^a}{d \phi_*}  \sim \mathcal{O} \left( \frac{\Delta}{\kappa_*} \right) \, .
\ee
This is a very reasonable situation to look into (our previous example is a particular case of this) as it correspond to those cases in which the main scale encoding the geometrical effects in the trajectory is its curvature. Then, if the non-vanishing eigenvalues of $C^a_{\phantom{a} b}$ are much larger than $\dot \phi_*^2 / \kappa^2$ we can neglect the first term in (\ref{eq-Delta-2}) and write
\be
 C^a_{\phantom{a} b} (\phi_*) \Delta^b \simeq  \frac{\dot \phi_*^2 N^a}{\kappa_*} \, . \label{eq-Delta-3}
\ee
Thus more generally $\Delta \simeq \dot \phi_*^2/(M^2 \kappa_{*})$ is indeed a good measure of the deviation from the true minimum.
Notice that in the case of a system with two scalar fields this is precisely the case.

\end{appendix}

\end{document}